\documentclass[twocolumn,showpacs,preprintnumbers,amsmath,amssymb,amsmath,amssymb,prb,floatfix,lengthcheck]{revtex4}

\usepackage{graphicx}
\usepackage{bm}
\usepackage{hyperref}
\usepackage{color}

\begin{document}

\title{An electrical probe for mechanical vibrations in suspended carbon nanotubes}

\author{N. Traverso Ziani$^{1}$, G. Piovano$^{1,2}$, F. Cavaliere$^{1,2}$ and M. Sassetti$^{1,2}$}
 \affiliation{$^1$ Dipartimento di Fisica, Universit\`a di Genova, Via Dodecaneso 33,
  16146, Genova, Italy.\\}
 \affiliation{$^2$ CNR-SPIN, Via Dodecaneso 33,
  16146, Genova, Italy.}
\date{\today}

\begin{abstract}
The transport properties of a suspended carbon nanotube probed by
means of a STM tip are investigated. A microscopic theory of the
coupling between electrons and mechanical vibrations is developed. It
predicts a {\em position-dependent} coupling constant, sizeable only
in the region where the vibron is located. This fact has profound
consequences on the transport properties, which allow to extract
information on the location and size of the vibrating portions of the
nanotube.
\end{abstract}

\maketitle

\section{Introduction}
\label{sec:intro}
Carbon nanotubes (CNTs)~\cite{scoperta} are extremely versatile
systems with metallic or semiconducting behavior depending on their wrapping
orientation.~\cite{saito,char} Deposing them on an insulating substrate and
tunnel-coupling it to biased electrodes it is possible to create a
single-electron transistor, in which the nanotube behaves as a quantum
dot.~\cite{bockrath,tans,cobden} Alternately one can embed a quantum
dot into the nanotube via geometrical defects or external gates, thus building a nanotube dot tunnel-coupled
to contacts.~\cite{postma}\\
\noindent Recent improvements in manipulation techniques
have allowed to \emph{suspend} nanotubes between
two contacts. In this case nanotubes behave as mechanical resonators,~\cite{hutt2} with possible
applications ranging from ultra-sensitive mass sensing to displacement
sensors.~\cite{stamp} Among the different mechanical
vibrations~\cite{zant} the radial breathing mode is  the highest in energy~\cite{leroy,leroy2,leroy3} followed by the
twist and the stretching ones. The latter have received a lot of
experimental attention ~\cite{saza,sapmaz,hutt,leturq} also in view of the peculiar
features induced on transport, such as negative differential
conductance~\cite{sapmaz} or Franck-Condon blockade.~\cite{leturq} 
Bending modes have usually energies lower than the experimental temperature requiring external AC
drivings.~\cite{hutt3,steele} 

\noindent Transport experiments have been employed to analyze the
structure of nanotubes exploiting a scanning tunneling
microscope (STM) tip.~\cite{lin} Effects such as spin-charge separation were
observed studying the differential conductance as a function of the
tip position.~\cite{lee} Superconducting probes have been used to extract the
non-equilibrium electron energy distribution function.~\cite{chen}
Also, the effects of chemical or magnetic impurities adsorbed along
the nanotube were considered.~\cite{clauss,furu,venema,lemay,ouyang,buchs}\\
\noindent Scanning tunnel microscopy experiments have also been
performed on \emph{suspended} nanotubes. In particular, it has been
shown how electrons injected from a tunnel microscope tip can excite, detect and control a specific
vibrational mode.~\cite{leroy,leroy2,leroy3}

\noindent Owing to their small waist (of the order of some
nm) nanotubes behave  as a one-dimensional interacting electronic system.~\cite{saito,char} Typically, correlated quantum systems are studied by means of numerical
techniques.~\cite{abedinpour,qian,degio,cavawi} However, due to
their inherently one-dimensional nature, carbon nanotubes are described  in terms of a  Luttinger model.~\cite{giamarchi,egger} In this context, transport from a
tunneling tip to a {\em static} nanotube has been recently
considered.~\cite{eggert,crep,leb,buchs,bercioux,bena}\\
\noindent The coupling between the electrons and 
vibrational modes has been extensively studied in
literature.~\cite{suzu,mart,penn,mahan,eros,flens,alves,izu} 
In most cases the simple Anderson-Holstein model has been
employed,~\cite{zazu,shen} in which the vibron couples only to the
\emph{total} charge neglecting the spatial modulation of the charge
density.  The Anderson-Holstein
interaction yields position-\emph{independent} Franck-Condon factors~\cite{franck,cond} with
visible effects in transport properties.~\cite{oppen,braig,haupt,merlo,piovano,koch}\\
Recently, a microscopic theory involving  the coupling with \emph{spatial fluctuations}
of the nanotube electronic density has also been developed~\cite{cava} in order to explain 
anomalous transport behaviors.~\cite{cava}

\noindent In this paper we investigate the possibility of creating an   
 {\em electrical probe} for the stretching vibrational modes
  of a suspended carbon nanotube, by means of a scanning
          tunnel microscope tip. Building on the theory outlined in
          Ref.~\onlinecite{cava}, we describe the coupling between
          vibrons and total charge as well as the spatial charge
          density modulations. This coupling gives rise to a
          position-dependent, electron-vibron coupling which strongly
          affects the transport properties. {\em Position-dependent}
          tunneling rates and conductance arise. This allows to obtain precise
informations about the vibrational mode of the nanotube.
\noindent Effects are visible in metallic nanotubes and are
more pronounced in semiconducting ones.

\noindent The paper is structured as follows. In Sec.~\ref{sec:modcnt}
a Luttinger liquid model for a carbon nanotube with open boundary conditions is introduced. In Secs.~\ref{sec:vib}-\ref{sec:vib1} the lattice
vibrations, the electron-vibron coupling and its diagonalization are
discussed. Section~\ref{sec:trans} is devoted to the transport
properties. Our results are illustrated and commented in
Sec.~\ref{sec:res}. Conclusions are drawn in Sec.~\ref{sec:concl}.

\section{Model and Methods}
\subsection{Modeling a carbon nanotube quantum dot}
\label{sec:modcnt}
\label{sec:model}
The electronic properties of the CNT are characterized by the {\em
  wrapping vector}
$\mathbf{w}_{n,m}=n\mathbf{a}_{+1}+m\mathbf{a}_{-1}$, where 
$\mathbf{a}_{\pm 1}$ represent the basis vectors of
the graphene lattice.~\cite{char} Due to
the wrapped nature of the system, the energy spectrum  is
composed of subbands corresponding to transverse excitations along the
waist of the tubule.  In a typical experiment only
the lowest-lying subband is occupied,~\cite{char,saito} giving a one dimensional character to the CNT.
In this regime \noindent both $n$-doped semiconducting CNTs (away from the band gap)
and metallic CNTs can be described in the low energy sector as Luttinger liquids with four
branches,~\cite{yo,grifoni,kleimann,cavaprl,spineffects} labeled by $\alpha=\pm 1$,
stemming from the two Dirac valleys of the graphene, and by $s=\pm 1$,
denoting the $z$ component of the electron spin (units $\hbar/2$).

\begin{figure}[h!]
\begin{center}
\includegraphics[width=8cm,keepaspectratio]{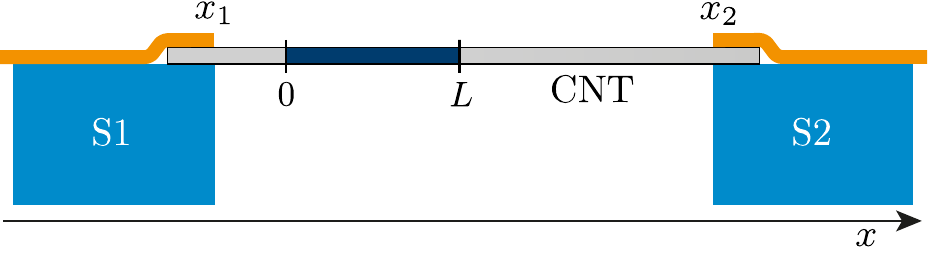}
\caption{(Color online) Schematic setup of a CNT, suspended
  between the two substrates S1 and S2 at positions $x_1$ and $x_2$.
  A quantum dot with ends at $x=0$ and $x=L$ is embedded in the CNT.}
\label{fig:fig1}
\end{center}
\end{figure}
\noindent The system under investigation is schematically depicted in
Fig.~\ref{fig:fig1}, it consists of a CNT, suspended between two
substrates at $x_1$ and $x_2$ and free to vibrate. Embedded in the CNT there is a quantum dot 
of length $L$ with ends at $x=0$ and $x=L$.  We assume $x_1\le 0$ and $x_2\ge L$ in order to mimic
a quantum dot inside the CNT, due to geometrical defect or external gates,
or to treat the CNT itself as the quantum dot.
In the following, we focus on the description of the quantum dot with
{\em open boundary conditions}.  

The bosonized hamiltonian is $\hat{H}=\sum_{j}\hat{H}_{j}$ with
\begin{equation}
\label{eq:hambos}
\hat{H}_{j}=\frac{1}{2}E_{j}\hat{N}_{j}^{2}+\sum_{q}\omega_{j}(q)\hat{b}_{j}^{\dagger}(q)\hat{b}_{j}(q)\, ,\quad\quad(\hbar=1)
\end{equation}
where $j\in\{\rho_{+},\rho_{-},\sigma_{+},\sigma_{-}\}$ are the four
linear combinations of states in the $\alpha,s$ branches that
diagonalize the Coulomb interaction. Here, $\hat{N}_{j}$
represent the zero modes counting the excess electrons in the $j$ sector
\begin{eqnarray}
\hat{N}_{\rho_{+}}=\sum_{\alpha,s}\hat{N}_{\alpha,s}\ &;&\ \hat{N}_{\rho_{-}}=\sum_{\alpha,s}\alpha\hat{N}_{\alpha,s}\,;\nonumber\\
\hat{N}_{\sigma_{+}}=\sum_{\alpha,s}s\hat{N}_{\alpha,s}\ &;&\ \hat{N}_{\sigma_{-}}=\sum_{\alpha,s}\alpha s\hat{N}_{\alpha,s}\, .\nonumber
\end{eqnarray}
The bosonic operators $\hat{b}_{j}(q)$ trigger collective
excitations of the electron system with momentum $q=\pi n/L$ with
$n\in\mathbb{N}^{*}$. They  are connected to the $\alpha,s$ modes
by a Bogoljubov transformation.~\cite{yo,grifoni} Note that the mode
$j=\rho_{+}$ represents the {\em total charge} of the system.

\noindent The collective modes propagates with velocities $v_{j}=v_{\mathrm
  F}/g_{j}$ with $g_{\rho_{+}}=g$ and $g_{j}=1$ $\forall j\neq\rho_{+}$. Here $g$
parameterizes the strength of electron scattering, with $g<1$ for
repulsive interactions. Note that only  the velocity of the total charge mode
is renormalized.
The corresponding energies are $\omega_{j}(q)=v_{j}q$ and $E_{j}=\pi v_{j}/4g_{j}L$.
\begin{figure}
\begin{center}
\includegraphics[width=5cm,keepaspectratio]{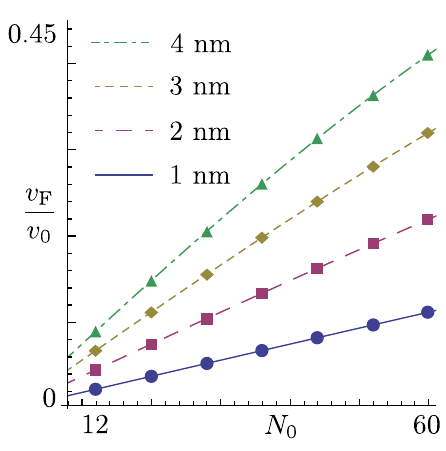}
\caption{(Color online)
Fermi velocity
$v_{\mathrm F}$ of a semiconducting CNT of length $L=400$ with
  $N_{0}$ electrons and different CNT waist lengths $|\mathbf{w}_{n,m}|$: blue (solid)
  1 nm - such as for (3,2), purple (dashed) 2 nm - such as for (5,4), yellow (dotted) 3 nm - such as for (8,6), green (dash-dotted) 4 nm - such as for (11,7).}
\label{fig:fig2}
\end{center}
\end{figure}	
\newline The value of the Fermi velocity $v_{\mathrm F}$ depends on the 
properties of the CNT. For a {\em metallic} CNT ($\!|n-m|/3\in\!\mathbb{N}$ )
the  dispersion relation is linear
with $v_{\mathrm{F}}=v_0=8\cdot10^{5}\ \mathrm{m}/\mathrm{s}$
around the degenerate point $\bar{k}$.
In the following, we will assume an effective $n$-doping with a shift of $E_{\mathrm{F}}$
towards higher energy values and  new Fermi points $k_{\mathrm{F}}^{(\pm)}=\bar{k}\pm
q_{\mathrm{F}}$ with $q_{\mathrm{F}}=E_{\mathrm{F}}/v_{\mathrm{F}}$.

\noindent In a {\em semiconducting} CNT the conduction and valence
bands are separated at momentum $\bar k'$  by the direct gap~\cite{char}
\begin{equation}
\Delta=\frac{4\pi v_{0}}{3|\mathbf{w}_{n,m}|}\, ,
\end{equation}
with $|\mathbf{w}_{n,m}|=a\sqrt{n^2+nm+m^2}$ the waist length of
the CNT. We will consider $n$-doped semiconducting nanotubes with
$E_{\mathrm{F}}>\Delta/2$, having chosen the energy reference to lie
in the middle of the band gap. Doping gives rise to two new Fermi
points $k_{\mathrm{F}}^{(\pm)}=\bar k'\pm q_{\mathrm{F}}$ where
$q_{\mathrm{F}}=2m^{*}\sqrt{E_{\mathrm{F}}-\Delta}$ with  $m^{*}=\Delta/2v_{0}^{2}$ the effective mass.~\cite{char} 
Around the new Fermi points, the Fermi velocity is 
\begin{equation}
v_{\mathrm{F}}=\frac{3|\mathbf{w}_{n,m}|N_{0}}{8L}\, ,
\end{equation}
with $N_{0}$ the total number of excess
electrons in the quantum dot.  The dependence of $v_{\mathrm{F}}$ as a
function of $N_{0}$ is shown in
Fig.~\ref{fig:fig2}. One observes a {\em lower} velocity with respect to the metallic case.

\noindent The electron field operator
$\hat{\Psi}_{s}(\mathbf{r})\equiv\hat{\Psi}_{s}(x,y)$ has to satisfy 
{\em open-boundaries} conditions
$\hat{\Psi}_{s}(0,y)=\hat{\Psi}_{s}(L,y)=0$.
\newline It can be written in the bosonized form
\begin{equation}
\label{eq:fieldopop}
\hat{\Psi}_{s}(\mathbf{r})=\sum_{r=\pm 1}\sum_{\alpha=\pm 1}f_{r,\alpha}(\mathbf{r})e^{irq_{\mathrm{F}}x}\hat{\psi}_{+1,r\alpha,s}(rx)
\end{equation}
in terms of the right-movers field operators 
\begin{eqnarray}
\hat{\psi}_{+1,\alpha,s}(x)&=&\frac{\hat{\eta}_{\alpha,s}}{\sqrt{2\pi\tilde{a}}}e^{-i\theta_{\alpha,s}}e^{i\frac{\pi
    x}{4L}\left(\hat{N}_{\rho_{+}}+\alpha
  \hat{N}_{\rho_{-}}+s\hat{N}_{\sigma_{+}}+\alpha s
  \hat{N}_{\sigma_{-}}\right)}\cdot\nonumber\\ &&e^{\frac{i}{2}\left[\hat{\phi}_{\rho_{+}}(x)+\alpha\hat{\phi}_{\rho_{-}}(x)+s\hat{\phi}_{\sigma_{+}}(x)+\alpha
    s \hat{\phi}_{\sigma_{-}}(x)\right]}\, ,\label{eq:fieldopdiag}
\end{eqnarray}
where
\begin{eqnarray}
\hat{\phi}_{j}(x)&=&\sum_{q}\sqrt{\frac{\pi}{qL}}\left\{\frac{1}{\sqrt{g_{j}}}\cos{(qx)}\left[\hat{b}_{j}(q)+\hat{b}_{j}^{\dagger}(q)\right]\nonumber\right.\\
&+&\left.i\sqrt{g_{j}}\sin(qx)\left[\hat{b}_{j}(q)-\hat{b}_{j}^{\dagger}(q)\right]\right\}\, .
\end{eqnarray}
Here $\hat{\eta}_{\alpha,s}$ are Majorana fermions,
$[\hat{\theta}_{\alpha,s},\hat{N}_{\alpha,s}]=i$, and $\tilde a$ is the length cutoff.
The functions  $f_{r,\alpha}(\mathbf{r})$ in
Eq.~(\ref{eq:fieldopop}) consist of a superposition of wavefunctions for $p_{z}$
orbitals, peaked around the positions of atoms in the CNT and
oscillating with a typical wave vector $K_{0}\propto a^{-1}$ where
$a\approx 2.5\cdot10^{-10}\ \text{m}$.~\cite{yo,grifoni} Their
specific form depends on the type of nanotube under
consideration and will be not discussed here.

\subsection{Lattice vibrations and electron-vibron coupling}
\label{sec:vib}
 We consider the case of a vibrating portion of the CNT (the vibron), of size
$\mathcal{L}$, located between $x_{v0}$ and $x_{v1}$. As confirmed in a
recent experiment,~\cite{cava} the vibrating part can be
different from the CNT dot. Thus, we will formulate the theory in this most general case.
We focus the description on the stretching mode. The $p$-th mode has  energy $\omega_{0}=p\pi v_{\mathrm s}/\mathcal{L}$ 
with  $v_{\mathrm s}\approx 2.4\cdot 10^{4}$ m/s the velocity of the stretching modes which is approximately
non-dispersive. In most experiments the fundamental mode with $p=1$ is
observed.~\cite{sapmaz,leturq} For these reasons, we will concentrate
to the case of {\em small} $p\leq 3$.
\noindent The $p$-th mode is described as a
harmonic oscillator
\begin{equation}
\label{eq:vibh0}
\hat{H}_{\mathrm v}=\frac{\hat{P_{0}}^2}{2M}+\frac{M\omega_{0}^2}{2}\hat{X}_{0}^2\, ,
\end{equation}
where $M=2\pi|\mathbf{w}_{n,m}|\mathcal{L}\rho_{0}$ is the vibron mass with
$\rho_{0}\approx6.7\cdot10^{-7}$ Kg/$\mathrm{m}^2$ the graphene
density and $\hat{X_{0}}$ is the amplitude operator of the strain field
\begin{equation}
\label{eq:strain}
\hat{u}_{p}(\mathbf{r})=\sqrt{2}\hat{X_{0}}\sin\left[p\frac{\pi(x-x_{v0})}{\mathcal{L}}\right]\, .
\end{equation}
The latter represents a standing
wave with momentum $\pm q_{0}$ with $q_{0}=p\pi/\mathcal{L}$.\\

\noindent The coupling between electrons and vibrations can be
microscopically derived starting from the tight-binding theory of a
distorted CNT lattice.~\cite{jishi,mahan} For the typical experimental situations one has
$\mathcal{L}\gtrsim 100$ nm then a continuum
elastic model is appropriate with~\cite{suzu}
\begin{equation}
\label{eq:ephint0}
H_{\mathrm{d-v}}=c\int\mathrm{d}\mathbf{r}\ \hat{\rho}(\mathbf{r})\partial_{x}\hat{u}_{p}(\mathbf{r})\,
,
\end{equation}
Here~\cite{suzu} $c\approx 30\ {\mathrm{eV}}$ and
$\hat{\rho}(\mathbf{r})=\sum_{s}\hat{\Psi}_{s}^{\dagger}(\mathbf{r})\hat{\Psi}_{s}(\mathbf{r})$
is the electronic density operator, with $\hat{\Psi}_{s}(\mathbf{r})$
given by Eq.~(\ref{eq:fieldopop}). It consists of two components: a
{\em long wavelength} part $\rho_{\mathrm{LW}}(\mathbf{r})$ and an
oscillatory contribution $\rho_{\mathrm{SW}}(\mathbf{r})$ fluctuating
on a length scale $K_{0}^{-1}\propto a$. This latter component,
does not make sizeable contributions  since $q_{0}\ll K_{0}$.
Under the realistic assumption of strongly localized atomic
orbitals~\cite{char,saito} with negligible overlapping, the
electron-vibron coupling becomes
\begin{equation}
\label{eq:ephint00}
H_{\mathrm{d-v}}=c\int_{x_{<}}^{x_{>}}\mathrm{d}x\ \hat{\rho}_{\mathrm{LW}}(x)\partial_{x}\hat{u}_{p}(x)\, ,
\end{equation}
with $x_{<}=\max{\{0,x_{v0}\}}$, $x_{>}=\min{\{x_{v1},L\}}$ and
\begin{equation}
\hat{\rho}_{\mathrm{LW}}=\frac{\hat{N}_{\rho_{+}}}{L}+\frac{1}{2\pi}\left[\partial_{x}\hat{\phi}_{\rho_{+}}(x)+x\to -x\right]\, ,
\end{equation}
written here directly in its bosonized form.
\subsection{Diagonalizing the electron-vibron coupling}
\label{sec:vib1}
The relevant terms of the electron-vibron coupling are
$\hat{h}=\hat{H}_{\rho_{+}}^{(0)}+\hat{H}^{(\rm{pl})}_{\rho_{+}}+\hat{H}_{\mathrm
  v}+\hat{H}_{\mathrm{d-v}}$ with
$H_{\rho_{+}}^{(0)}=E_{\rho_{+}}\hat{N}_{\rho_{+}}^{2}/2$ and
$\hat{H}_{\rho_{+}}^{(\mathrm{pl})}=\sum_{q}\omega_{\rho_{+}}(q)\hat{b}_{\rho_{+}}(q)\hat{b}_{\rho_{+}}(q)$. Introducing
$\hat{B}_{\mu}=i\hat{b}_{\rho_{+}}(\pi \mu/L)$ and
$\sqrt{2\omega_{\rho_{+}}(\pi\mu/L)}\hat{X}_{\mu}=\hat{B}_{\mu}+\hat{B}_{\mu}^{\dagger}$
we have
\begin{eqnarray}
\hat{h}&=&\frac{1}{2}E_{\rho_{+}}\hat{N}_{\rho_{+}}^{2}+\frac{\hat{P}_{0}^{2}}{2M}+\frac{M\omega_{0}^{2}}{2}\hat{X}_{0}^{2}+\sum_{\mu\geq1}\left(\frac{\hat{P}_{\mu}^{2}}{2}+\omega_{\mu}^{2}\frac{\hat{X}_{\mu}^{2}}{2}\right)\nonumber\\ &+&\sqrt{M}C_{0}\hat{X}_{0}\hat{N}_{\rho_{+}}+\sqrt{M}\hat{X}_{0}\sum_{\mu\geq1}C_{\mu}\hat{X}_{\mu}\,
,\label{eq:twocouplings}
\end{eqnarray}
with $[\hat{X}_{\mu},\hat{P}_{\nu}]=i\delta_{\mu,\nu}$,
$\omega_{\mu}=\mu\omega_{1}/g$, $\omega_{1}=\pi v_{\rm F}/L$. In terms of these new variables the density operator is 
\begin{equation}
\hat{\rho}_{\mathrm{LW}}(x)=\frac{\hat{N}_{\rho_{+}}}{L}+\sqrt{\frac{2\pi
    v_{\rm F}}{L^3}}\sum_{\mu\geq 1}\mu\cos\left(\frac{\pi \mu
  x}{L}\right)\hat{X}_{\mu}\, .
\end{equation}
The last term in  Eq.~(\ref{eq:twocouplings}) describes a {\em
 central harmonic oscillator} (vibron) linearly coupled to a infinity of harmonic
oscillators (plasmon modes of the dot).  Note that, for
reasonable experimental parameters and considering the lowest
stretching modes, one always has $\omega_{\mu}>\omega_{0}$ both for
metallic and semiconducting CNTs. Additionally, the vibron is also
coupled to the total average charge $\hat{N}_{\rho_{+}}$ on the quantum dot
in analogy to the Anderson-Holstein model.  The
coupling coefficients are
\begin{equation}
\label{eq:coefs}
C_{0}=\sqrt{2}\lambda_{\mathrm{m}}\omega_{0}^{3/2}J_{0}\quad;\quad C_{\mu\geq 1}=2\lambda_{\mathrm{m}}\omega_{0}^{3/2}\sqrt{\omega_{1}}J_{\mu}
\end{equation}
where ($\kappa\ge0$)
\begin{equation}
J_{\kappa}=\frac{1}{L}\int_{x_{<}}^{x_{>}}{\mathrm d}x\ \cos\left[\frac{\kappa\pi x}{L}\right]\cos\left[\frac{p\pi}{\mathcal{L}}(x-x_{v0})\right]\, ,
\end{equation}
and
\begin{equation}
\label{eq:lambdam}
\lambda_{\mathrm m}=\frac{c}{v_{\mathrm s}\sqrt{\rho_{0}\pi|\mathbf{w}_{n,m}|v_{\mathrm s}}}\, .
\end{equation}
 Taking as a
reference~\cite{char,saito} a waist length of about 2 nm, one has
$\lambda_{\mathrm m}\approx 2$. Note that a recent
experiment~\cite{bolotin} reports a larger $c$ which leads to a larger
$\lambda_{\mathrm{m}}$. One finds
\begin{equation}
J_{\kappa}=J_{\kappa}^{(0)}+J_{\kappa}^{(-)}\theta(-0^{+}-x_{v0})+J_{\kappa}^{(+)}\theta(x_{1}-L-0^{+})\, ,
\end{equation}
with $\theta(x)$ the Heavyside step function and
\begin{eqnarray}
\label{eq:Jk}
J_{\kappa}^{(0)}&=&\frac{\kappa\delta^2\left\{\sin\left(\!\pi \kappa \xi_{0}\right)-(-1)^{p}\sin\left[\pi \kappa (\xi_{0}+\delta)\right]\right\}}{\pi(p^2-\kappa^2\delta^2)}\, ,\nonumber\\
J_{\kappa}^{(-)}&=&\frac{p\delta\sin\left(\!\pi p \xi_{0}/\delta\right)-\kappa\delta^2\sin\left(\pi \kappa \xi_{0}\right)}{\pi(p^2-\kappa^2\delta^2)}\, ,\nonumber\\
J_{\kappa}^{(+)}&=&\frac{(-1)^{p}\kappa\delta^2\sin\left[\pi \kappa (\xi_{0}+\delta)\right]}{\pi(p^2-\kappa^2\delta^2)}\nonumber\\
&-&\frac{p\delta(-1)^{\kappa}\sin\left[\pi p\delta (1-\xi_{0})/\delta\right]}{\pi(p^2-\kappa^2\delta^2)}\, ,\nonumber
\end{eqnarray}
where $\delta={\mathcal L}/L$ and $\xi_{0}=x_{v0}/L$.\\ 

\noindent Let us now comment the general features of the
coupling depending on the relative size and position of dot and
vibron. When the vibron is {\em much larger} than the dot ($\delta\gg 1$) and the
latter is embedded into it one has
$J_{\kappa}\approx\delta_{\kappa,0}$, namely  in the large vibron
limit the coupling to the charge density fluctuations {\em vanishes}
and only the conventional Anderson-Holstein coupling survives. In this
limit, the electron-vibron coupling reduces to the standard
form~\cite{piovanoprb}
$\omega_{0}\lambda_{\mathrm{m}}\ell_{0}^{-1}\hat{X}_{0}\hat{N}_{\rho_{+}}$
with $\ell_{0}^{-1}=\sqrt{M\omega_{0}}$, and the coupling constant is $\lambda_{\mathrm{m}}$.\\
\noindent The most interesting case occurs when the vibron is {\em
  smaller} than the dot and embedded into it. In this regime, one
finds $J_{0}\equiv 0$, while $J_{\kappa\ge1}\neq 0$. This fact signals
a {\em radical departure} from the Anderson-Holstein model: the
coupling between electrons and vibrons occurs {\em only} via the
spatial fluctuations of the electron density.

\noindent We now turn to the diagonalization of the electron-vibron
coupling. The terms linear in $\hat{X}_{\mu}$ in
Eq.~(\ref{eq:twocouplings}) can be exactly
diagonalized,~\cite{ullersma} leading to
\begin{eqnarray}
\hat{h}&=&\frac{1}{2}E_{\rho_{+}}\hat{N}_{\rho_{+}}^{2}+\sum_{\mu\geq0}\left(\frac{\hat{\bar{P}}_{\mu}^{2}}{2}+\Omega_{\mu}^{2}\frac{\hat{\bar{X}}_{\mu}^{2}}{2}\right)\nonumber\\
&+&\sqrt{M}C_{0}\left(\sum_{\nu\geq 0}k_{0\nu}\hat{\bar{X}}_{\nu}\right)\hat{N}_{\rho_{+}}\, ,\label{eq:dopolaron}
\end{eqnarray}
with
\begin{equation}
\left(\hat{X}_{\mu},\hat{P}_{\mu}\right)=\sum_{\nu\geq0}k_{\mu\nu}\left(\hat{\bar{X}}_{\nu},\hat{\bar{P}}_{\nu}\right)\, .\label{eq:trasformazione}
\end{equation}
In order to diagonalize the term $\propto\hat{N}_{\rho_{+}}$ a
Lang-Firsov canonical transformation is used
\begin{equation}
\hat{\mathcal{U}}=e^{-i\hat{N}_{\rho_{+}}\sum_{\nu\geq 0}\eta_{\nu}\hat{\bar{P}}_{\nu}}\, ,\label{eq:poltra}
\end{equation}
with
$\eta_{\nu}=\sqrt{M}C_{0}k_{0\nu}/(\Omega_{\nu}^{2}\sqrt{M})$. This
leads to a shift
\begin{equation}
\hat{\bar{X}}_{\nu}\to\hat{\bar{X}}_{\nu}-\eta_{\nu}\hat{N}_{\rho_{+}}\, ,
\end{equation}
which finally casts the hamiltonian into the diagonal form
\begin{equation}
\hat{h}=\frac{1}{2}\left(E_{\rho_{+}}-\Delta E_{\rho_{+}}\right)\hat{N}_{\rho_{+}}^{2}+\sum_{\mu\geq0}\left(\frac{\hat{\bar{P}}_{\mu}^{2}}{2}+\Omega_{\mu}^{2}\frac{\hat{\bar{X}}_{\mu}^{2}}{2}\right)\, ,
\end{equation}
with $\Delta E_{\rho_{+}}=C_{0}^{2}\sum_{\nu\geq 0}(k_{0\nu}^2/\Omega_{\nu}^{2})$.\\

\noindent The energies $\Omega_{\mu}$ of the new eigenmodes are the
roots of the secular equation
\begin{equation}
z^{2}=\omega_{0}^{2}+\sum_{\nu\geq1}\frac{C_{\nu}^{2}}{z^{2}-\nu^2\omega_{1}^{2}}\, ;\label{eq:omega}
\end{equation}
with
\begin{eqnarray}
k_{\mu\nu}&=&\frac{C_{\mu}}{\Omega_{\nu}^{2}-\mu^2\omega_{1}^{2}}k_{0\nu}\quad({\mathrm{with}}\ \mu\geq 1)\, ,\label{eq:kmunu}\\
k_{0\nu}&=&\left[1+\sum_{\mu\geq1}\frac{C_{\mu}^{2}}{\left(\Omega_{\nu}^{2}-\mu^2\omega_{1}^{2}\right)^2}\right]^{-1/2}\, .\label{eq:k0nu}
\end{eqnarray}
\begin{figure}[h!]
\begin{center}
\includegraphics[width=8cm,keepaspectratio]{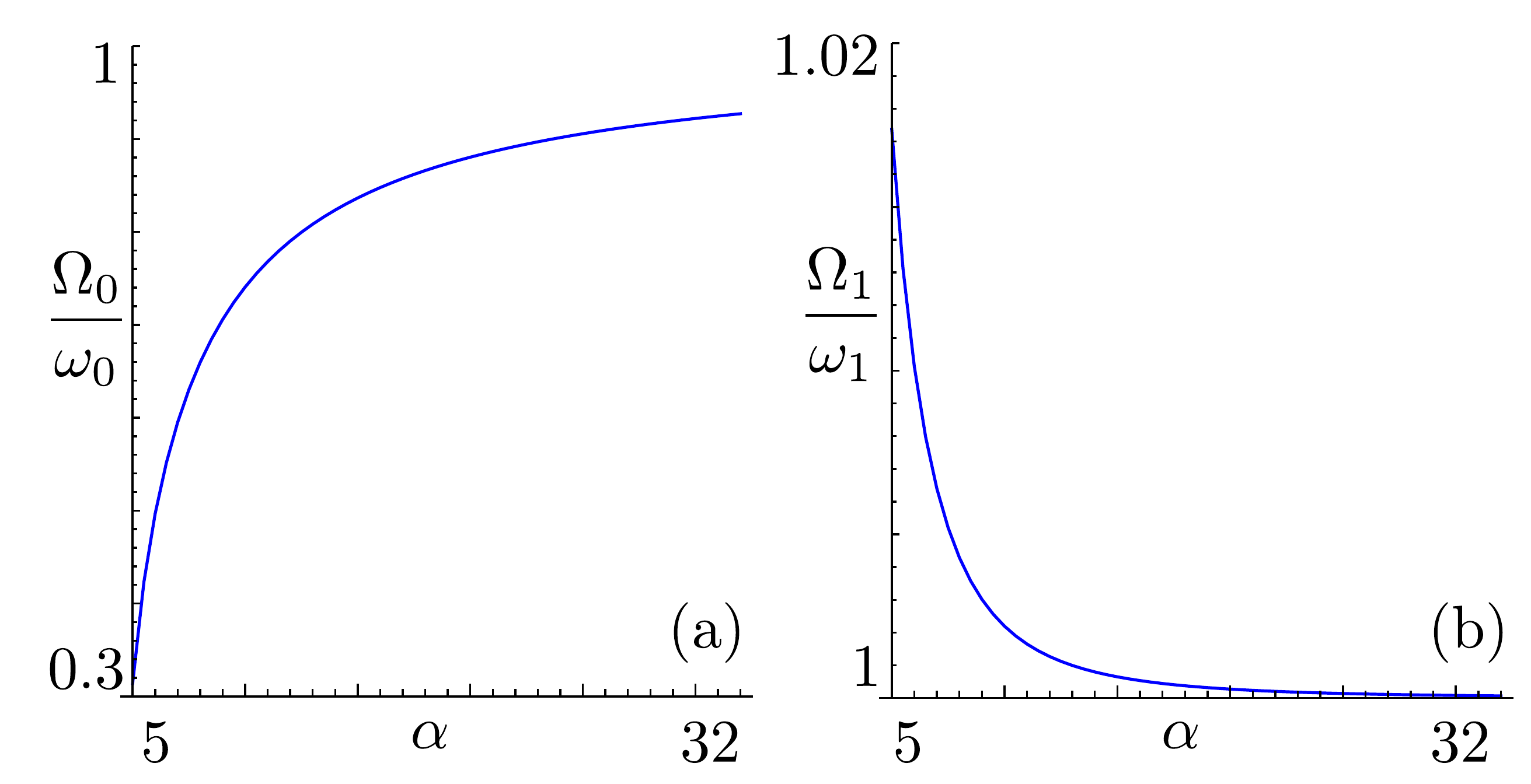}
\caption{Energy of the lowest eigenmode. (a) Plot of
  $\Omega_{0}/\omega_{0}$ as a function of $\alpha$; (b) Plot of
  $\Omega_{1}/\omega_{1}$ as a function of $\alpha$. In all figures,
  $\delta=1$, $x_{v0}=0$, $p=1$, $g=1$ and
  $\lambda_{\mathrm m}=2$.}
\label{fig:fig3}
\end{center}
\end{figure}
\noindent As can be clearly seen, the modes
with $\mu\geq 1$ have energies $\Omega_{\mu}\gtrsim\mu\omega_{1}$ and
represent blue-shifted dressed plasmons. The lowest-lying solution, on
the other hand, has $\Omega_{0}<\omega_{0}$ and represents a dressed
vibronic mode red-shifted by the electron-vibron interaction. By
inspecting Eq.~(\ref{eq:omega}) one always obtains a real solution for
$\Omega_{\mu}$ with $\mu\geq 1$. On the other
hand, the existence of a real solution for $\Omega_{0}$ requires
$\omega_{0}^{2}{\omega_{1}^{2}}>\sum_{\mu\geq
  1}C_{\mu}^{2}/\mu^2$.
When this condition is not fulfilled, the Wentzel-Bardeen instability
occurs.~\cite{mart} In our calculations we have always checked that for realistic
parameters the system does not exhibit this instability.
\noindent The energy of the dressed vibronic mode is very sensitive to
the ratio $\alpha=v_{\mathrm  F}/v_{\mathrm s}$ between the Fermi and the sound velocity.
While for $\alpha=32$, corresponding to the case
of a metallic CNT, one has $\Omega_{0}\approx\omega_{0}$,
for lower values of $\alpha$, typical of a semiconducting CNT,
a suppression of $\Omega_{0}$ occurs, see
Fig.~\ref{fig:fig3}(a). Note that the dressed plasmons are almost
insensitive to the ratio $\alpha$ (cf.  Fig.~\ref{fig:fig3}(b)).\\

\noindent The transformations in Eq.~(\ref{eq:trasformazione}) and
(\ref{eq:poltra}) affects the electronic field operator of
Eq.~(\ref{eq:fieldopdiag}). Up to an irrelevant phase constant, the
field $\hat{\phi}_{\rho_{+}}(x)$ is 
\begin{equation}
\hat{\phi}_{\rho_{+}}(x)=\sum_{\mu\geq
  0}\alpha_{\mu}(x)\hat{\bar{X}}_{\mu}+\beta_{\mu}(x)\hat{\bar{P}}_{\mu},
\end{equation}
where
\begin{eqnarray}
\label{eq:alpha}
\alpha_{\mu}(x)&=&\sqrt{2\omega_{1}}\sum_{\nu\geq1}k_{\nu \mu}\sin\left(\frac{\pi\nu x}{L}\right)\\
\label{eq:beta}
\beta_{\mu}(x)&=&\sqrt{M}\eta_{\mu}+\sqrt{\frac{2}{\omega_{1}}}\sum_{\nu\geq1}\frac{k_{\nu \mu}}{\nu}\cos\left(\frac{\pi\nu x}{L}\right)\, .
\end{eqnarray}
\subsection{Modeling transport}
\label{sec:trans}
\begin{figure}[h!]
\begin{center}
\includegraphics[width=8cm,keepaspectratio]{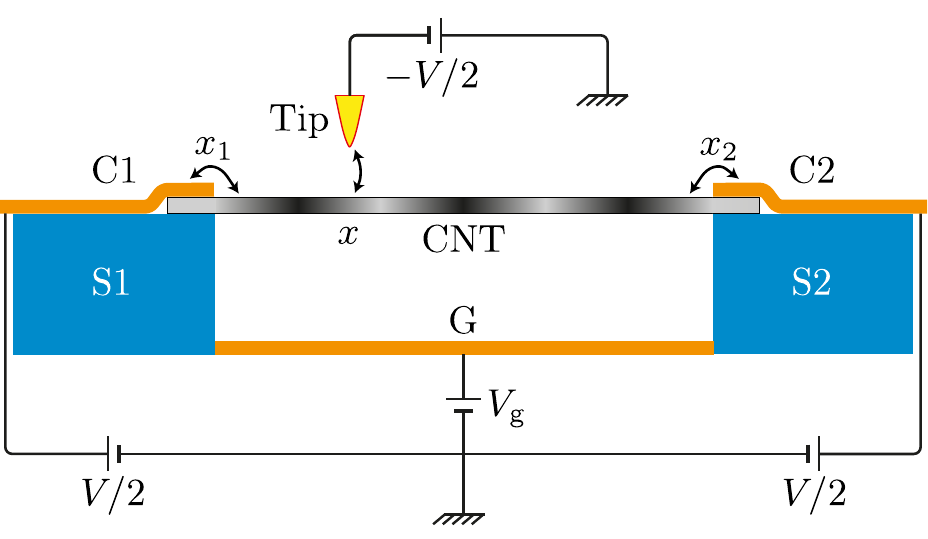}
\caption{Schematic setup of a STM transport experiment performed on a
  suspended CNT. The quantum dot is the nanotube itself, biased with respect to the tip and
  two lateral contacts.  Electrons can flow through the system
  via the STM  tip at position ($x$) and at
  the contacts (${x}_{1,2}$). A back gate allows to tune the
  effective charge on the dot.}
\label{fig:fig3-trans}
\end{center}
\end{figure}
Our task is to model an STM tunneling tip  on a 
CNT.~\cite{leroy,leroy2,leroy3} The setup is sketched in
Fig.~\ref{fig:fig3-trans} and is composed of a STM tip and two lateral
contacts tunnel-coupled to a suspended CNT. Tip and contacts are
biased in such a way that, for $V>0$, electrons flow
from the tip to the contacts through the CNT. Additionally, a back
gate is capacitively coupled to tune the effective charge on the
CNT-dot. From now on, we will focus on the most interesting
regime, namely that of a quantum dot along all the CNT with a vibron imbedded into it. 
The hamiltonian for the vibrating CNT is then 
\begin{equation}
  H_{\mathrm{CNT}}=\frac{1}{2}E_{\rho_{+}}\left(\hat{N}_{\rho_{+}}-N_{\mathrm
    g}\right)^{2}+\sum_{\mu\geq
    0}\frac{\hat{\bar{P}}_{\mu}^{2}}{2}+\frac{\Omega_{\mu}^{2}}{2}\hat{\bar{X}}_{\mu}^{2}+\sum_{j\neq\rho_{+}}\hat{H}_{j}\,
  ,\nonumber
\end{equation}
where $N_{\mathrm g}$ represents the charge induced by the back gate voltage $V_{\mathrm g}$. The dot is laterally coupled to the
two Fermi contacts, via the tunneling hamiltonian~\cite{cava}
\begin{equation}
\hat{H}_{\mathrm{T,L}}=t_{0}\sum_{j=1,2}\sum_{\alpha,s,q}\hat{\psi}_{+1,\alpha,s}(x_{j})\hat{c}_{j,s}(q)+\mathrm{h.c.}\,,
\end{equation}
where $t_{0}$ is the tunneling amplitude, $\hat{c}_{j,s}(q)$ are the
operators for an electron with momentum $q$ and spin $s$ in the
non-interacting lead $j$ and ${x}_{1}=0$, ${x}_{2}=L$ are the
position of the tunneling contacts.\\
\noindent The STM tip is modeled as a semi-infinite Fermi contact,
placed above the CNT at a position $x$ along it. The tunnel coupling
is expressed in terms of the Fermi field operator for the forward
modes of the tip~\cite{buchs,bercioux} $\hat{\psi}_{s,\mathrm{F}}(z)$
($z$ is the coordinate along the tip with $z=0$ at the vertex)
\begin{equation}
\!\hat{H}_{\mathrm{T,T}}\!\!=\!\!\left[\int_{0}^{L}\mathrm{d}y\ \tau_{0}(y-x)\sum_{\alpha,s}\hat{\psi}_{+1,\alpha,s}^{\dagger}(y)\right]\hat{\psi}_{s,\mathrm{F}}(0^{+})+\mathrm{h.c.}\,
.\nonumber
\end{equation}
The function $\tau_{0}(x)=\tau_{0}\varphi(x)$ describes the geometry
of the tip, with $0\leq\varphi(x)\leq 1$ peaked around
$x=0$, where $\varphi(0)=1$. In the following we will assume a typical
tip, with an effective width of a few atomic cells of the CNT. A
voltage $-V/2$ is applied to the STM tip, while the lateral contacts
are kept at the same voltage $V/2$. Voltage drops are assumed to occur
symmetrically on the CNT. More general potential distributions do not
affect the results at a qualitative level.

\noindent We will consider the sequential tunneling regime, treating
the tunnel couplings to the lowest perturbative order.
Since we are interested into the low-energy transport regime ($e|V|k_{\rm{B}}T\approx\Omega_{0}$) we
disregard the dynamics of the modes $\mu\geq 1$. Furthermore, we
will consider the relevant situation of a damped vibronic mode, with a
thermal equilibrium distribution at temperature $T$.\\

\noindent The eigenstates of the suspended CNT can be expressed by $|\{N_{\alpha,s}\}\rangle$ specifying
the distribution of excess electrons in the channel $\alpha$ with spin
$s$. We will consider the resonance between the state of a closed shell with $N_{0}=4\kappa$,  ($\kappa$ integer)
electrons and zero excess charges,  denoted as $|0\rangle=|0,0,0,0\rangle$, and $N_{0}+1$ electrons in the state $\alpha$,$s$.  Note
that no qualitative difference in our results would occur, for a
different value of $N_{0}$.  There are four states with $N_{0}+1$,
electrons, denoted by $|\alpha,s\rangle$, each with one extra electron
in the state $\alpha,s$.  They are all degenerate, with energy
$E_{\rho_{+}}(1-N_{\mathrm g})^{2}/2+3\omega_{1}/8$.
\noindent We set up a master equation for the reduced density matrix,
obtained tracing out the leads and vibron degrees of freedom - thus
neglecting coherences among vibrational states.~\cite{piovanoprb,brandesnew,yarnew} Upon
the assumption of a STM tip width of some unit cells, coherence
effects between states $|\alpha,s\rangle$ and $|\alpha',s\rangle$
($\alpha\neq\alpha'$) are vanishing. Coherence between different spin
states is also absent in view of the absence of spin correlations in
the contacts. Therefore, the master equation reduces to a standard
rate equation for the occupation probability of the quantum dot
$P_{N_{0}}(t)=P_{|0\rangle}(t)\,;\quad P_{N_{0}+1}(t)=\sum_{\alpha,s}P_{|\alpha,s\rangle}(t)$.
The steady-state current is then written, to lowest order, in terms  of tunneling rates
\begin{equation}
\label{eq:current}
I(x)=e\frac{\Gamma^{(\rm{C})}_{\mathrm{out}}\Gamma_{\mathrm{in}}^{(\mathrm{T})}(x)-\Gamma^{(\rm{C})}_{\mathrm{in}}\Gamma_{\mathrm{out}}^{(\mathrm{T})}(x)}{\Gamma_{\mathrm{in}}(x)+\Gamma_{\mathrm{out}}(x)}\,,
\end{equation}
with 
\begin{equation}
\Gamma_{\mathrm{in/out}}(x)=\Gamma_{\mathrm{in/out}}^{(\mathrm T)}(x)+\Gamma_{\mathrm{in/out}}^{(\rm{C})}
\end{equation}
and
\begin{eqnarray}
&&\!\!\Gamma_{\mathrm{in}}^{(\mathrm{T})}(x)=\sum_{\alpha,s}\Gamma_{|0\rangle\to|\alpha,s\rangle}^{(\mathrm{T})}(x)\ \ ;\ \ \Gamma_{\mathrm{out}}^{(\mathrm{T})}(x)=\Gamma_{|\alpha,s\rangle\to|0\rangle}^{(\mathrm{T})}(x)\, ,\nonumber\\
&&\!\!\Gamma_{\mathrm{in}}^{(\mathrm{C})}=\sum_{j}\sum_{\alpha,s}\Gamma_{|0\rangle\to|\alpha,s\rangle}^{(j)}\ \ ;\ \ \Gamma_{\mathrm{out}}^{(\mathrm{C})}=\sum_{j}\Gamma_{|\alpha,s\rangle\to|0\rangle}^{(j)}\nonumber\, ,
\end{eqnarray}
the rate associated to the tip $(\rm T)$ and to the contacts $(\rm C)$.

\noindent We quote here explicitly the expressions for the tunnel-in processes, the tunnel-out rates are similar. 

The lateral contact and tip rates are respectively  
\begin{eqnarray}
\label{eq:gamcon}
\!\!\!\!\!\!\!\!\!\Gamma_{|0\rangle\to|\alpha,s\rangle}^{(j)}&=&\Gamma_{0}\sum_{l\geq 0}B_{l}({x}_{j})f\left(\Delta E+l\Omega_{0}+eV/2\right),\\
\label{eq:gamtip}
\!\!\!\!\!\!\!\!\!\Gamma_{|0\rangle\to|\alpha,s\rangle}^{(\mathrm{T})}(x)&=&\Gamma_{0}^{(\mathrm{T})}\sum_{l\geq 0}B_{l}(x)f\left(\Delta E+l\Omega_{0}-eV/2\right).
\end{eqnarray}
Here, $\Gamma_{0}=2\pi\nu_{0}|t_{0}|^2$,  $\Gamma^{(\rm{T})}_{0}=2\pi\nu_{0}|\tau_{0}|^2$, $\nu_{0}$ is the leads density of state and $f(E)$ the Fermi function.
The rates are then a superposition of Fermi functions at energies 
\begin{equation}
\label{eq:DELTAE}
\Delta E=E_{\rho_{+}}\left(\frac{1}{2}-N_{\mathrm
  g}\right)+\frac{3}{8}\omega_{1}\, ,
\end{equation}
shifted by the energy $l\Omega_{0}$, representing the contribution of a transport channel exciting $l$ vibron quanta. 
The weights $B_{l}(x)$ considered in the regime $k_{\mathrm
  B}T\ll\Omega_{0}$ are~\cite{braig}
\begin{equation}
\label{eq:weights}
B_{l}(x)=\frac{\lambda^{2l}(x)}{l!}e^{-\lambda^{2}(x)}\, ,
\end{equation}
where
\begin{equation}
\label{eq:lambdaofx}
\lambda^{2}(x)=\frac{1}{2\Omega_{0}}\alpha_{0}^{2}(x)+\frac{\Omega_{0}}{2}\beta_{0}^{2}(x)
\end{equation}
represents the {\em local electron-vibron couplingstrength},~\cite{cava}
see Eqns.~(\ref{eq:alpha},\ref{eq:beta}).
This is in sharp contrast to the case of the
Anderson-Holstein model,~\cite{braig,koch,haupt,merlo} appropriate for
large vibrons. In that case, as discussed above, coupling to the
density fluctuations would vanish leading to $\alpha_0(x)=\beta_0(x)=0$
and to a coupling simply given by $\lambda_{\mathrm{max}}$ which is clearly
{\em independent} of the tunneling position.

\section{Results}
\label{sec:res}
\subsection{Local electron-vibron coupling}
Let us analyze  in details the space dependence of the electron-vibron coupling $\lambda(x)$
which determines the current behavior.  
The coupling $\lambda(x)$ depends both on geometrical parameters $x_{v0}$ (the
vibron origin), $\delta={\mathcal L}/L$ (vibron length) and on 
physical ones $\alpha=v_{\mathrm F}/v_{\mathrm s}$, $\lambda_{\mathrm
  m}$ defined in Eq.~(\ref{eq:lambdam}) and the electronic
interaction parameter $g$.

\noindent We remind that the parameter $\alpha$ is affected by the metallic or
semiconducting nature of the CNT. In a metallic CNT one finds
$\alpha=32$. On the other hand, for a semiconducting CNT lower values of $\alpha$ are possible (see later).
\begin{figure}[h!]
\begin{center}
\includegraphics[width=8cm,keepaspectratio]{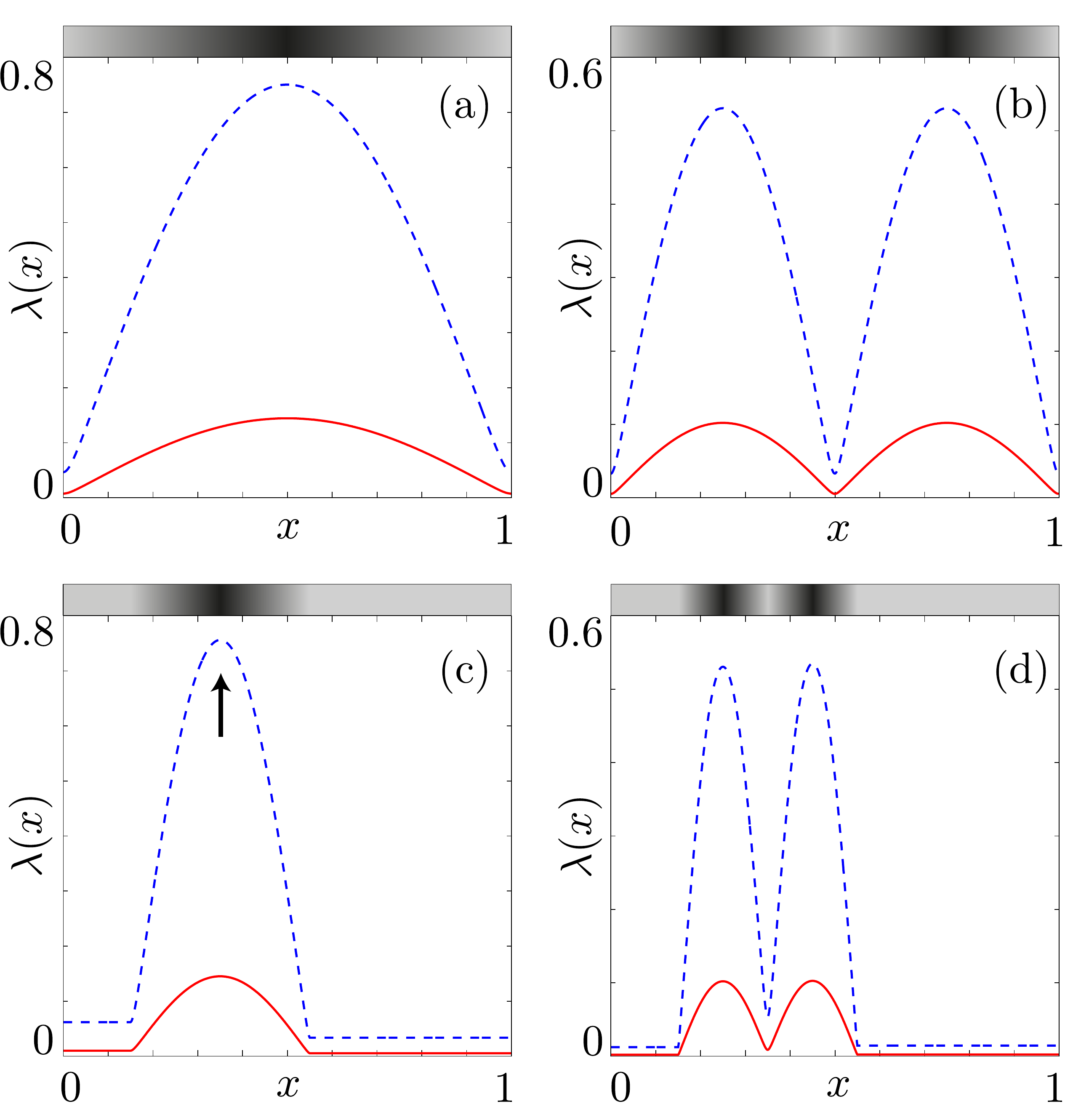}
\caption{(Color online) Local electron-vibron coupling $\lambda(x)$ as
  a function of the tip position $x$  for: (a) $\delta=1$, $x_{v0}=0$,
  $p=1$; (b) as above but $p=2$; (c) $\delta=0.4$, $x_{v0}=0.15L$,
  $p=1$; (d) as above but $p=2$. In all panels red (solid) lines
  denote a metallic CNT with $\alpha=32$, blue (dashed) lines denote a
  semiconducting CNT with $\alpha=5$. Other parameters:
  $\lambda_{\mathrm m}=2$ and $g=1$. The shaded plots on top of the
  panels depict the amplitude of the strain field $\hat{u}_{p}(x)$. The arrow in Panel (c) denotes the
  tip position for the conductance shown in Fig.~\ref{fig:fig7}.}
\label{fig:fig4}
\end{center}
\end{figure}
Figure~\ref{fig:fig4} shows $\lambda(x)$ for different vibron configurations and CNT types.
It can be seen that the electron-vibron coupling strength
for a metallic CNT (solid red lines) is smaller than that for a
semiconducting CNT. Indeed, for a semiconducting CNT, the velocities of the electronic and
vibronic subsystems are closer, which implies a more favorable
interplay between them. 
In the rest of the paper, we will choose $\alpha=5$ to model a
semiconducting CNT and $\alpha=32$ for the metallic one.\\
\noindent The amplitude of the electron-vibron coupling is maximal in
the region where the strain field is maximum. Indeed, $\lambda(x)$
closely follows the amplitude of $\hat{u}_{p}(x)$, which is sketched
on top of the panels of Fig.~\ref{fig:fig4}. For $\delta<1$, this
implies a particularly sizeable $\lambda(x)$ only in the region where
the vibron sits, see Figs.~\ref{fig:fig4}(c,d).
\noindent Coupling to higher vibronic modes produces more
oscillations, as can be seen in Figs.~\ref{fig:fig4}(b,d). It also
makes the electron-vibron coupling strength weaker.
\begin{figure}[h!]
\begin{center}
\includegraphics[width=8cm,keepaspectratio]{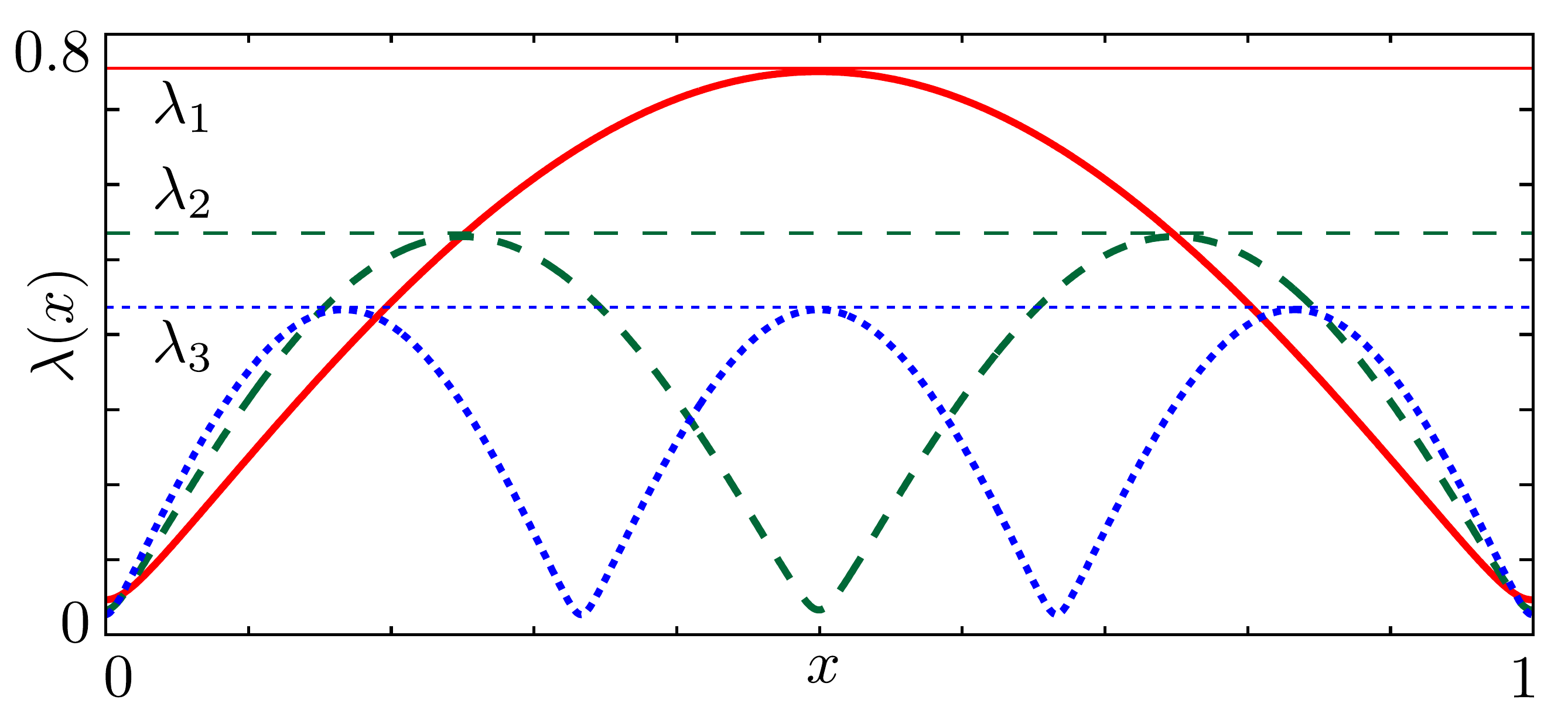}
\caption{(Color online) Local electron vibron coupling $\lambda(x)$ as
  a function of $x$ for $\delta=1$, $x_{v0}=0$ and different vibron
  modes: red (solid) $p=1$, green (dashed) $p=2$, blue (dotted)
  $p=3$. The thin lines are the maxima of
  $\lambda(x)$. Other parameters: $\alpha=5$, $\lambda_{\mathrm m}=2$, $g=1$.}
\label{fig:fig5}
\end{center}
\end{figure}
Figure~\ref{fig:fig5} shows the comparison between the
first three vibronic modes. The intensity of the
electron-vibron coupling strength {\em decreases} with the increasing
order of the vibronic mode. Denoting $\lambda_{p}$ the maximum of
$\lambda(x)$ for the $p$ mode, we find
$\lambda_{p}=\lambda_{1}/\sqrt{p}$.
\begin{figure}[h!]
\begin{center}
\includegraphics[width=8cm,keepaspectratio]{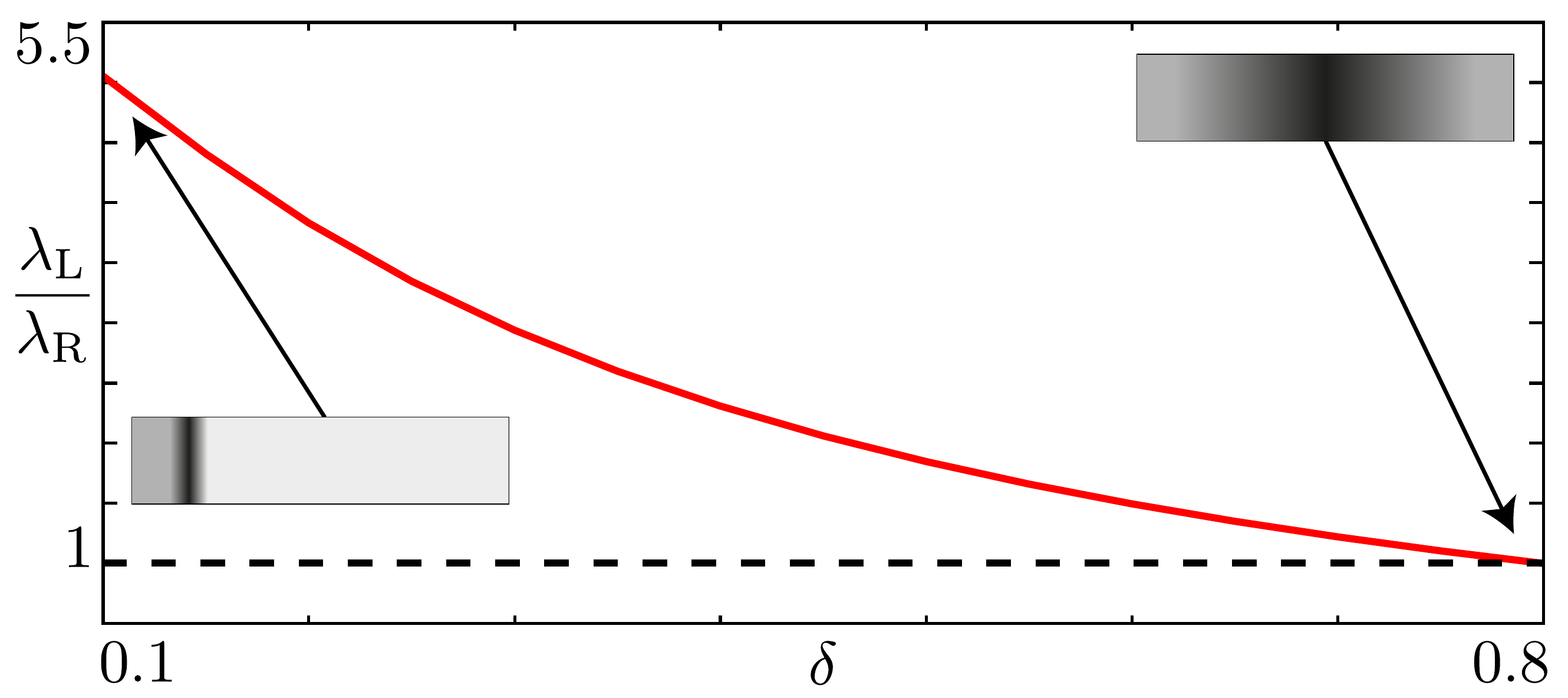}
\caption{(Color online) Ratio of the coupling strengths
  $\lambda_{\mathrm L}/\lambda_{\mathrm R}$ (see text) as a function of $\delta$ for a vibron with origin at
  $x_{v0}=0.1L$ and $p=1$. The shaded plots schematically depict the
  amplitude of the strain field of the vibronic mode $\hat{u}_{p}(x)$.
  Other parameters: $\alpha=5$, $\lambda_{\mathrm m}=2$ and $g=1$.}
\label{fig:fig6}
\end{center}
\end{figure}
We now briefly comment on the value of the electron-vibron coupling at
the position of the tunneling barriers $\lambda_{\mathrm
  L}\equiv\lambda(0)$ and $\lambda_{\mathrm R}\equiv\lambda(L)$, which  
govern the lateral tunneling rates. Figure~\ref{fig:fig6} shows the
ratio $\lambda_{\mathrm L}/\lambda_{\mathrm R}$ as a function of
$\delta$ for a vibron with origin at $x_{v0}=0.1L$. At small values of
$\delta$ the vibron is asymmetrically located near the left
tunnel barrier. As a consequence, $\lambda_{\mathrm
  L}>\lambda_{\mathrm R}$. This mechanism is at the origin of the
systematic suppression of conductance traces in a recent
experiment.~\cite{cava} For increasing $\delta$ the situation evolves
towards a more symmetric setup and indeed for $\delta=0.8$,
corresponding to a symmetric vibron with respect to the CNT, one
recovers $\lambda_{\mathrm L}=\lambda_{\mathrm R}$.
\subsection{Transport properties}
\begin{figure}[h!]
\begin{center}
\includegraphics[width=6cm,keepaspectratio]{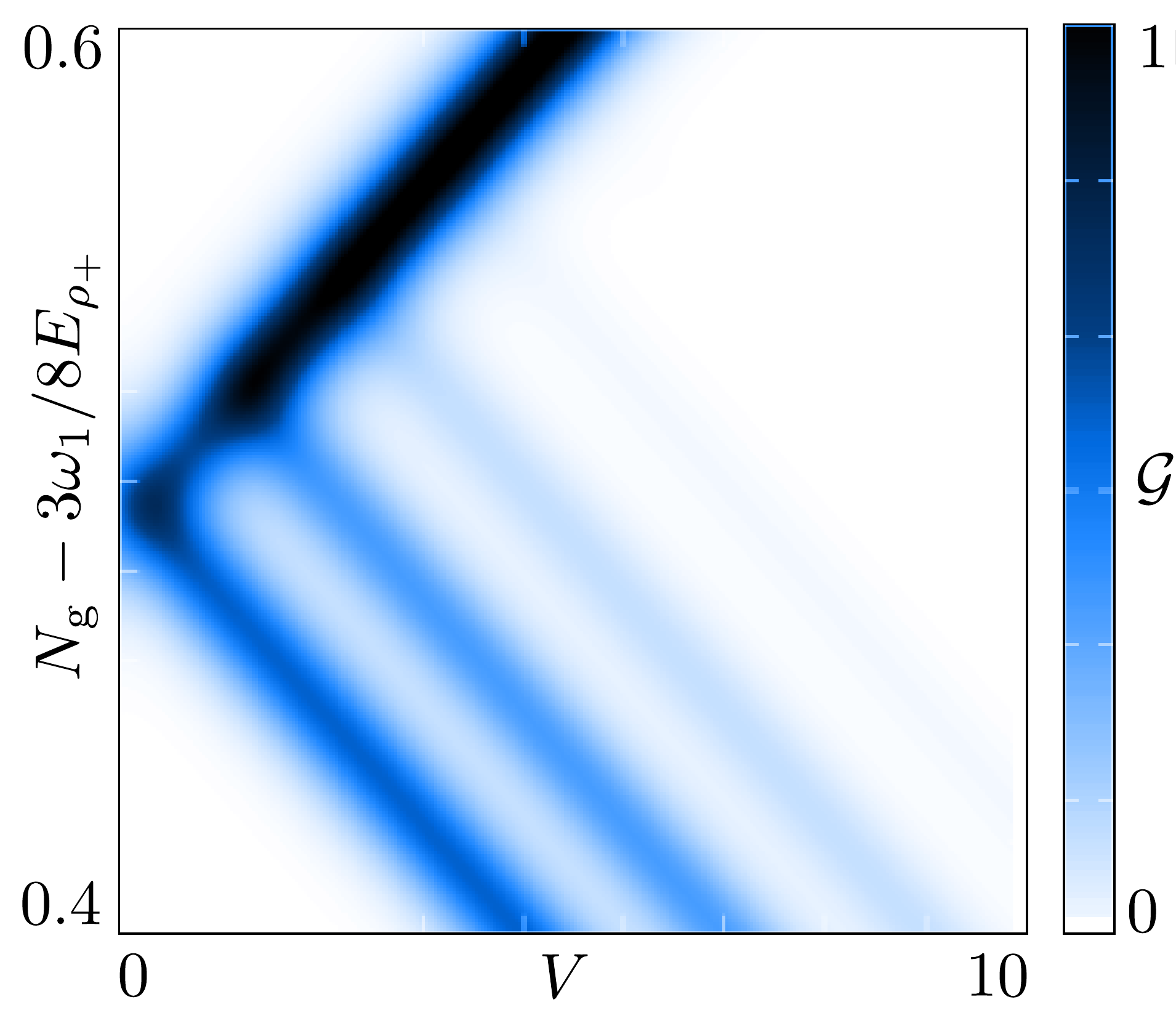}
\caption{(Color online) Semiconducting CNT. Differential conductance
  $\mathcal{G}$ (units $e^2\Gamma_{0}^{(\mathrm{T})}/\Omega_{0}$) as a function of
  $N_{\mathrm g}$ and $V$ (units $\Omega_{0}/e$) for a vibron
  originating at $x_{v0}=0.15L$ with $\delta=0.4$, $p=1$ and a tip at
  $x=0.35L$. Other parameters: $\lambda_{\mathrm m}=2$, $\alpha=5$,
  $g=1$, $k_{\mathrm B}T=0.15\,\Omega_{0}$ and
  $A=\Gamma_{0}/\Gamma_{0}^{(\mathrm{T})}=100$.}
\label{fig:fig7}
\end{center}
\end{figure}

Figure~\ref{fig:fig7} shows the density plot of the
differential conductance $\mathcal{G}=\partial I/\partial V$ in the
$(V,N_{\mathrm g})$ plane for the situation depicted in
Fig.~\ref{fig:fig4}(c), semiconducting case.
\noindent The large white areas at small $V$ are the Coulomb blockade
regions where transport is interdicted and the CNT is occupied by $N_0$ or $N_0+1$ electrons.
Within the transport region,
delimited by the two most intense conductance traces, a series of
equally spaced lines are clearly visible. They correspond
to the excitation of the vibronic mode at energy $\Omega_{0}$.
We consider different  tunneling amplitudes through the tip and the lateral contacts,
introducing the  {\em asymmetry} 
$A=\Gamma_{0}/\Gamma_{0}^{(\mathrm{T})}$.\\

\noindent  We will concentrate the discussion on the regime
$k_{\mathrm{B}}T<\Omega_{0}$, quoting  for simplicity 
analytical expressions for $A\gg1$ only, realistic in an STM
experiment. Exploiting  the fact that $\lambda_{\mathrm{L,R}}\ll1$ we assume
$B_{l}(x_{j})\approx\delta_{l,0}$.
 In the {\em linear} regime ($V\to 0$) the conductance is  then
\begin{equation}
\mathcal{G}_{\mathrm{lin}}\approx\frac{2\beta
  e^2\Gamma_{0}^{({\mathrm T})}B_{0}(x)}{\cosh\left[\frac{\beta\Delta
      E}{2}-\frac{1}{2}\ln(4)\right]\cosh\left[\frac{\beta\Delta E}{2}\right]}\,
,\label{eq:condlinasympt}
\end{equation}
with $\Delta E$ in Eq.~(\ref{eq:DELTAE}). The logarithmic
factor $\ln(4)$ stems from the fourfold degeneracy of the state with
$N_{0}+1$ electrons.
The amplitude of the linear conductance
is modulated by the factor
\begin{equation}
B_{0}(x)=e^{-\lambda^2(x)}\, .
\end{equation}
Therefore, the conductance is {\em suppressed} in the region
where the electron-vibron coupling is
large.

\noindent In the {\em nonlinear} regime ($V>k_{\mathrm
  B}T$) one finds
\begin{equation}
\mathcal{G}_{\mathrm{nonlin}}\approx\frac{\beta
  e^2\Gamma_{0}}{2}\sum_{l\geq0}\frac{B_{l}(x)}{\cosh\left[\beta\frac{\Delta
    E+l\Omega_{0}-eV/2}{2}\right]}\, .\label{eq:condasympt}
\end{equation}
\noindent Equation~(\ref{eq:condasympt}) represents a fan of
equally-spaced conductance peak lines located at
$E_{\rho_{+}}(1-2N_{\mathrm
  g})+3\omega_{1}/8E_{\rho_{+}}+l\Omega_{0}-eV/2=0$, thus with
negative slope in the $(V,N_{\mathrm g})$ plane, see
Fig.~\ref{fig:fig7}. They originate by the tunneling from the STM
tip to the CNT, triggering vibronic excitations. We note that, because
of the smallness of $\lambda_{\mathrm{L,R}}$, the triggering process
due to the tunneling on the contacts barriers is strongly suppressed
(since $B_{l}(x_{j})\approx\delta_{l,0}$) with a corresponding
absence of conductance lines with positive slope. Each of the above
peaks is weighted by $B_{l}(x)$ which in turn conveys informations on
$\lambda(x)$. This is particularly clear for the $l$-th ($l\geq 1$)
{\em nonlinear} conductance peak. Indeed, for $\lambda(x)<1$ (which is
the case considered in our calculations, see Fig.~\ref{fig:fig4}) one
finds that $B_{l}(x)$ is a {\em monotonically increasing} function of
$\lambda(x)$ - see Eq.~(\ref{eq:weights}) - which implies an increase
of the conductance for increasing coupling strength. This fact allows
to directly map the spatial modulations of the nonlinear differential
conductance into modulations of $\lambda(x)$.

\noindent To be more specific, let us consider the resonance case
\begin{equation}
\Delta E=0\Longleftrightarrow N_{\mathrm g}=\frac{1}{2}+\frac{3\omega_{1}}{8E_{\rho_{+}}}
\end{equation}
and study $\mathcal{G}$.
\begin{figure}[h!]
\begin{center}
\includegraphics[width=8cm,keepaspectratio]{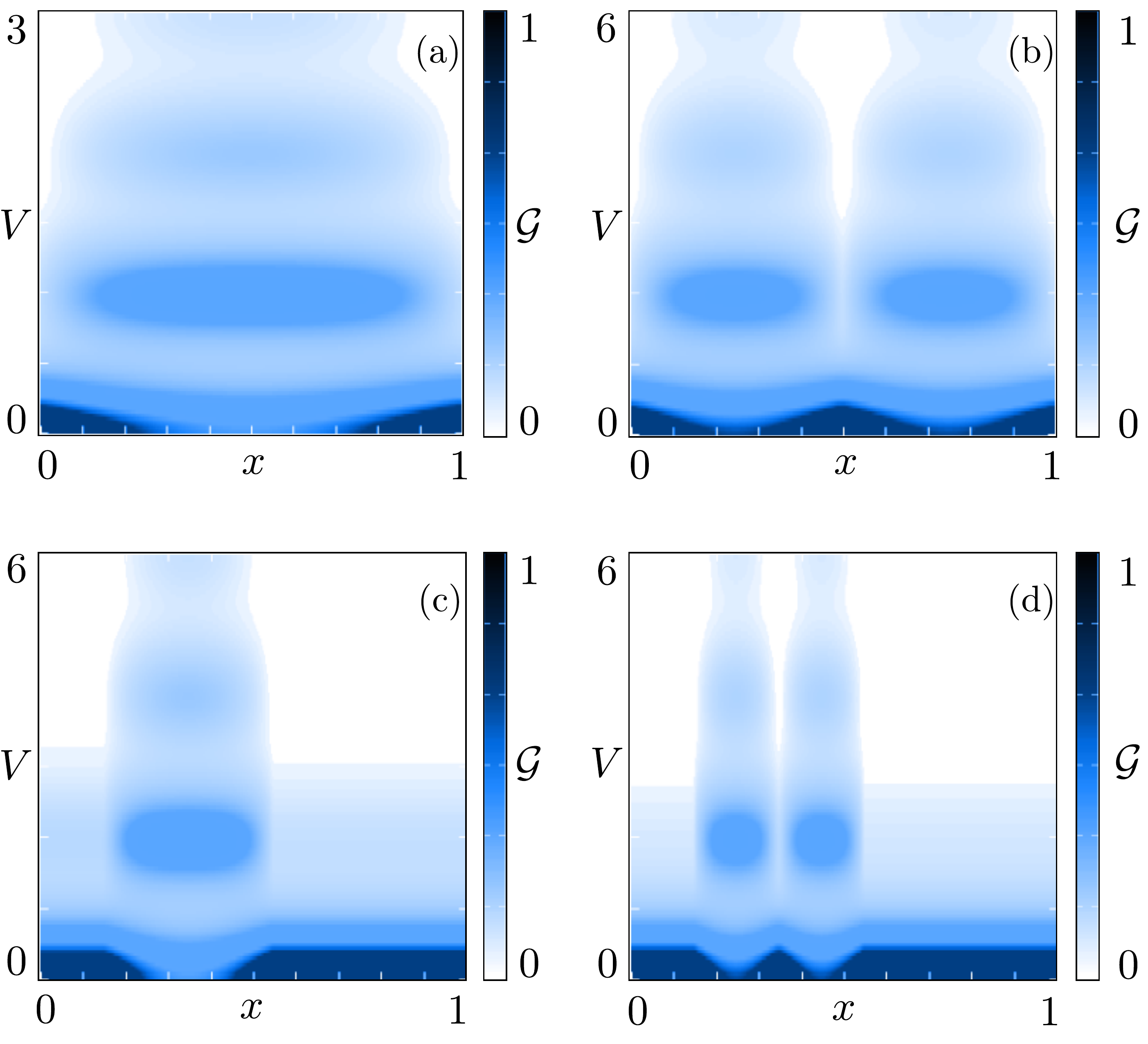}
\caption{(Color online) Semiconducting CNT: Plot of $\mathcal{G}$
  (units $e^2\Gamma_{0}^{(\mathrm{T})}/\Omega_{0}$) as a function of
  the tip position $x$ and bias voltage $V$ (units $\Omega_{0}/e$) at
  resonance $N_{\mathrm g}=(1/2)+(3\omega_{1}/8E_{\rho_{+}})$ and (a)
  $\delta=1$, $x_{v0}=0$ and $p=1$; (b) same as in (a) but for $p=2$;
  (c) $\delta=0.4$, $x_{v0}=0.15L$ and $p=1$; (d) same as in (c) but
  for $p=2$. Other parameters: $\lambda_{\mathrm m}=2$, $\alpha=5$,
  $g=1$, $k_{\mathrm B}T=0.15\,\Omega_{0}$ and $A=100$.}
\label{fig:fig8}
\end{center}
\end{figure}
\noindent As is clear by inspecting Figs.~\ref{fig:fig8}(a-d), the
differential conductance $\mathcal{G}$ exhibits position-dependent
modulations in close agreement to the behavior of $\lambda(x)$.  The
most striking features show up indeed near the vibron position, where
$\mathcal{G}$ is suppressed for $V\approx 0$, and is enhanced for
$eV\approx 2l\Omega_{0}$ ($l=1,2,\ldots$), see Fig.~\ref{fig:fig4}.

\begin{figure}[h!]
\begin{center}
\includegraphics[width=8cm,keepaspectratio]{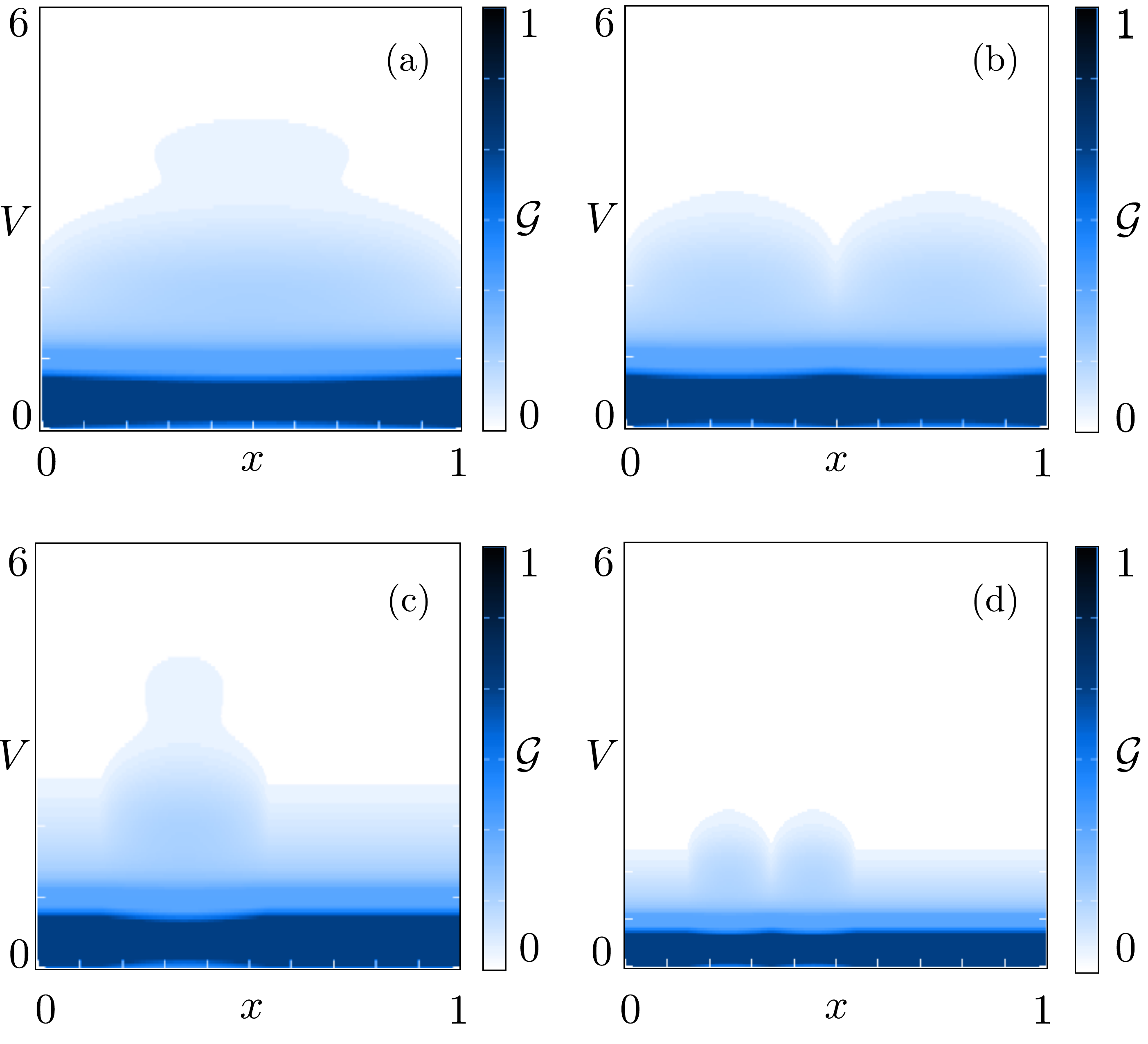}
\caption{(Color online) Metallic CNT: Plot of $\mathcal{G}$ (units
  $e^2\Gamma_{0}^{(\mathrm{T})}/\Omega_{0}$) as a function of the tip position $x$
  and bias voltage $V$ (units $\Omega_{0}/e$) at resonance $N_{\mathrm
    g}=(1/2)+(3\omega_{1}/8E_{\rho_{+}})$ and (a) $\delta=1$,
  $x_{v0}=0$ and $p=1$; (b) same as in (a) but for $p=2$; (c)
  $\delta=0.4$, $x_0=0.15L$ and $p=1$; (d) same as in (c) but for $p=2$. Other
  parameters: $\lambda_{\mathrm m}=2$, $\alpha=32$, $g=1$,
  $k_{\mathrm B}T=0.15\,\Omega_{0}$ and $A=100$.}
\label{fig:fig9}
\end{center}
\end{figure}

\noindent For the case of a metallic CNT, shown in
Fig.~\ref{fig:fig9}, the spatial modulations of the conductance are
less pronounced due to the decreased intensity of the electron-vibron
coupling with respect to the semiconducting case.

\begin{figure}[h!]
\begin{center}
\includegraphics[width=8cm,keepaspectratio]{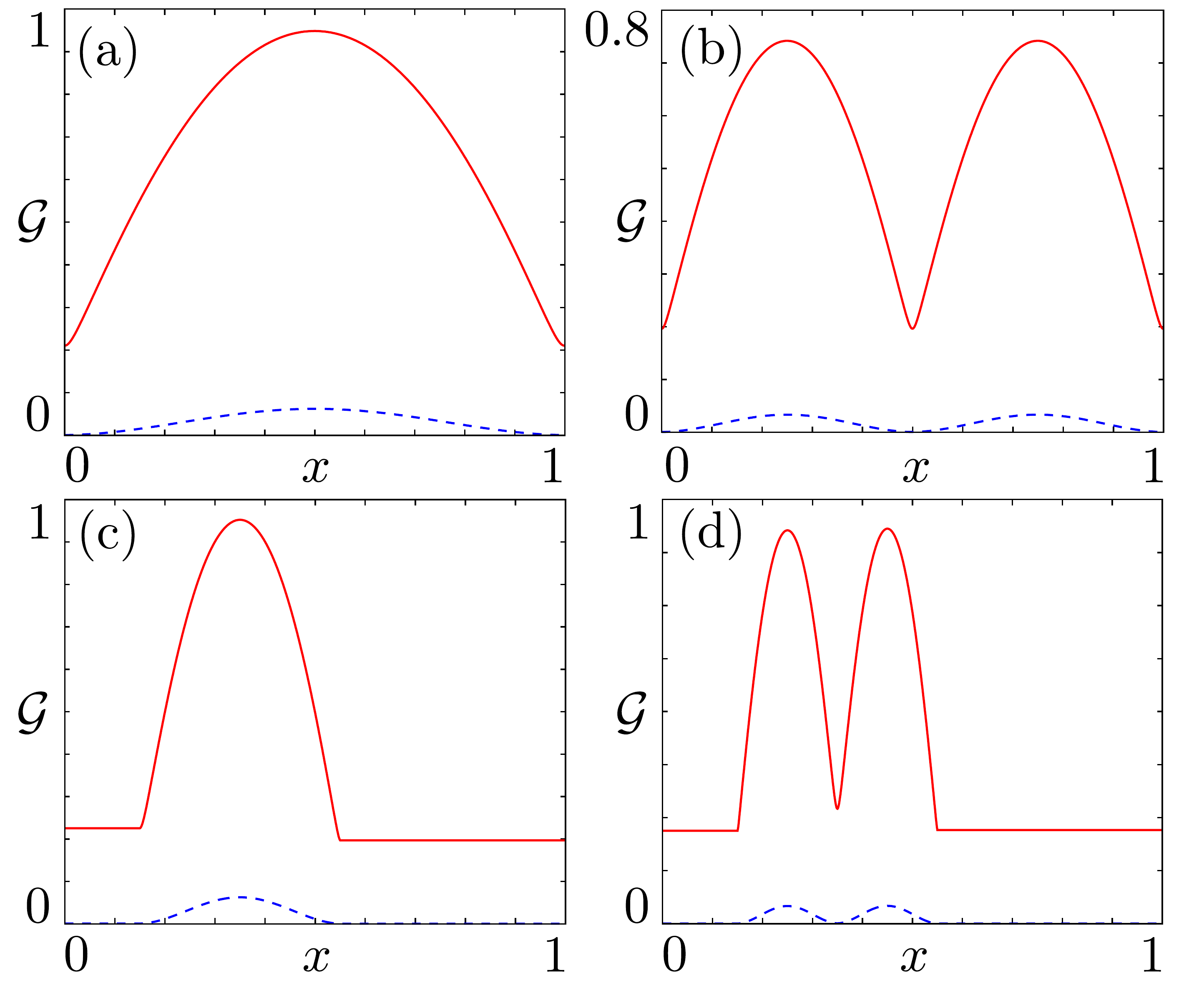}
\caption{(Color online) Differential conductance $\mathcal{G}$ (units
  $e^2\Gamma_{0}^{(\mathrm{T})}/\Omega_{0}$) as a function of the tip position $x$
  for $N_{\mathrm g}=(1/2)+(3\omega_{1}/8E_{\rho_{+}})$ and
  $eV=2\Omega_{0}$ (red solid line) or $eV=4\Omega_{0}$ (blue dashed
  line) for  a metallic CNT: (a) $\delta=1$, $x_{v0}=0$ and
  $p=1$; (b) same as in (a) but for $p=2$; (c) $\delta=0.4$,
  $x_{v0}=0.15L$ and $p=1$; (d) same as in (c) but for $p=2$. Other
  parameters: $\lambda_{\mathrm m}=2$, $\alpha=32$, $g=1$,
  $k_{\mathrm B}T=0.15\,\Omega_{0}$ and $A=100$.}
\label{fig:fig10}
\end{center}
\end{figure}

\noindent The close resemblance of the spatial modulations of the non
linear $\mathcal{G}$ with $\lambda(x)$ is supported studying the
conductance at fixed bias shown in Fig.~\ref{fig:fig10} for
$eV=2\Omega_{0}$ and $eV=4\Omega_{0}$.
Clearly $\mathcal{G}$ is {\em enhanced} where $\lambda(x)$ is
large. This confirms that position-resolved conductance maps are
source of valuable informations about the intensity of the strain
field along the CNT and consequently on the location and size of the
vibron mode.

\subsection{Interaction and asymmetry effects}
We close by briefly commenting about the role of electron-electron
interactions and that of the barriers asymmetry.
\begin{figure}[h!]
\begin{center}
\includegraphics[width=8cm,keepaspectratio]{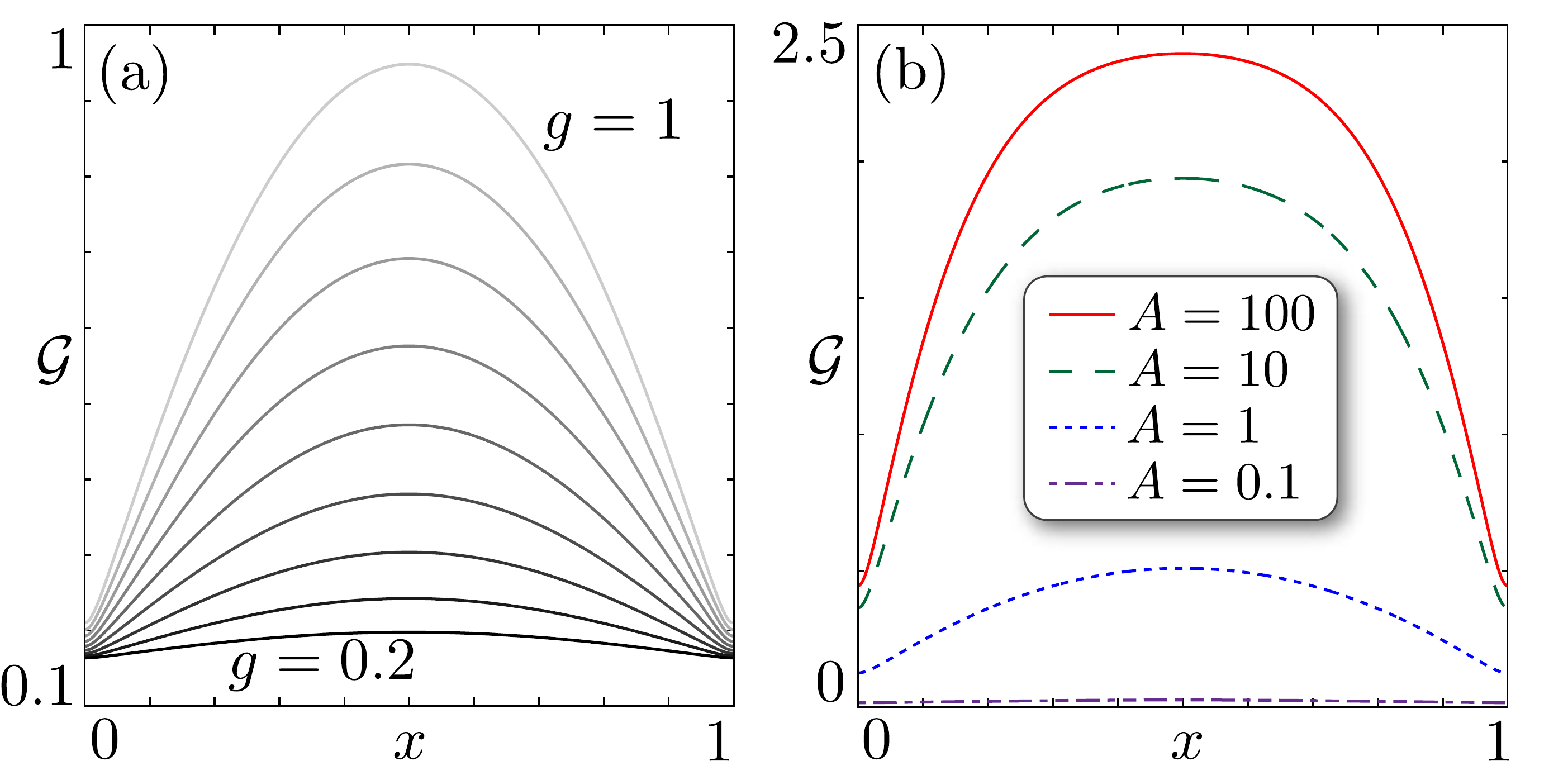}
\caption{(Color online) (a) Conductance $\mathcal{G}$ (units
  $e^2\Gamma_{0}^{(\mathrm{T})}/\Omega_{0}$) as a function of the tip
  position $x$ for decreasing values of $g$ from $g=1$
  (noninteracting, lightest gray) to $g=0.2$ (strongly interacting,
  black); (b) Plot of $\mathcal{G}$ for different asymmetries (see key) and $g=1$. Other
  parameters:  $\delta=1$, $x_{v0}=0$,  $p=1$, $\lambda_{\mathrm m}=2$, $\alpha=5$, $k_{\mathrm
    B}T=0.15\,\Omega_{0}$ and $eV=2\Omega_{0}$.}
\label{fig:fig11}
\end{center}
\end{figure}

\noindent Figure~\ref{fig:fig11}(a) shows the position-resolved
nonlinear conductance at $eV=2\Omega_{0}$ for increasing Coulomb
interaction strength (decreasing values of $g$). Clearly, conductance
is suppressed in turns, signaling the suppression of $\lambda(x)$. This
fact can be explained in terms of an increase of the velocity of the
charged mode $v_{\rho_{+}}=v_{\rm{F}}/g$ which induces an {\em
  effective} parameter $\alpha$ {\em higher} than that of the
noninteracting case.

\noindent Concerning the role of the asymmetry between the tip rate
and the leads rate, Fig.~\ref{fig:fig11}(b) shows that the
amplitude of the conductance modulations is decreased when making the
tunnel barriers more symmetric with a collapse when reversing the
asymmetry $A<1$ (i.e. making the contacts more opaque than the STM
tip). This can be understood by observing that for $A\gg 1$ the
current is dominated by the slowest barrier which is the STM one. This
implies that the space-dependent tunneling rate can be efficiently
probed in this regime.\\

\section{Conclusions}
\label{sec:concl}
In this paper we have shown how transport measurements, performed with
a scanning tunnel microscope tip on a suspended carbon nanotube, can
bring information about its vibrational stretching dynamics. This theory predicts a {\em
  position-dependent} coupling constant, which is larger in the region
where the vibron is located.

\noindent The position-dependent coupling constant strongly affects
the tunneling rate of electrons through the tip and has relevant
consequences in the transport spectra. In
particular, we showed that conductance maps in the
linear and nonlinear regime, obtained sweeping the tip along the
nanotube, are closely connected to the local coupling constant and
allow to localize the position and size of the vibron. Effects can be
more pronounced in semiconducting nanotubes due to the reduced Fermi
velocity which matches more closely the speed of the vibrational
mode. The role of electronic interactions and of the asymmetry between
tip and metal contact tunnel barriers have also been addressed.

\noindent This work could inspire a new class of experiments which aim
at studying the vibrational degrees of freedom of a nanotube by means
of electrical measurements.\\

\noindent\textit{Acknowledgments.} The authors acknowledge stimulating
discussions with V. Cataudella, E. Mariani, A. Nocera, E. Paladino and
C. Stampfer. Financial support by the EU-FP7 via ITN-2008-234970 NANOCTM
is also gratefully acknowledged.


\begin{thebibliography}{99}
\bibitem{scoperta} S. Iijima, Nature (London) \textbf{354}, 56 (1991).
\bibitem{char} J.-C. Charlier, X. Blase, and S. Roche,
  Rev. Mod. Phys. \textbf{79}, 677 (2007).
\bibitem{saito} R. Saito, G. Dresselhaus, and M. S. Dresselhaus,
  \textit{`Physical Properties of Carbon Nanotubes'}, Imperial College
  Press (1998).
\bibitem{cobden} D. H. Cobden and J. Nyg\aa rd, Phys. Rev. Lett. {\bf
  89}, 046803 (2002).
\bibitem{bockrath} M. Bockrath, D. H. Cobden, P. L. McEuen,
  N. G. Chopra, A. Zettl, A. Thess, and R. E. Smalley, Science {\bf
    725}, 1922 (1997).
\bibitem{tans} S. J. Tans, M. H. Devoret, H. Dai, A. Thess,
  R. E. Smalley, L. J. Geerligs, and C. Dekker, Nature {\bf 386}, 474
  (1997).
\bibitem{postma} H. W. C. Postma, T. Teepen, Z. Yao, M. Grifoni, and
  C. Dekker, Science \textbf{293}, 76 (2001).
\bibitem{hutt2} A. K. H\"uttel, G. A. Steele, B. Witkamp, M. Poot,
  L. P. Kouwenhoven, and H. S. J. van der Zant, Nano Lett. \textbf{9},
  2547 (2009).
\bibitem{stamp} C. Stampfer, A. Jungen, R. Linderman, D. Obergfell,
  S. Roth, and C. Hierold, Nano Lett. \textbf{6}, 1449 (2006).
\bibitem{zant} M. Poot and H. S. J. van der Zant, to appear on
  Phys. Rep., arXiv:1106.2060v1.
\bibitem{leroy} B. J. Leroy, S. G. Lemay, J. Kong, and C. Dekker,
  Appl. Phys. Lett. \textbf{84}, 4280 (2004).
\bibitem{leroy2} B. J. LeRoy, S. G. Lemay, J. Kong, and C. Dekker,
  Nature \textbf{432}, 371 (2004).
\bibitem{leroy3} B. J. LeRoy, I. Heller, V. K. Pahilwani, C. Dekker,
  and S. G. Lemay, Nano Lett. \textbf{7}, 2937 (2007).
\bibitem{saza} V. Sazonova, Y. Yaish, H. \"Ust\"unel, D. Roundy,
  T. A. Arias, and P. L. McEuen, Nature \textbf{431}, 284 (2004).
\bibitem{sapmaz} S. Sapmaz, P. Jarillo-Herrero, Y. M. Blanter,
  C. Dekker, and H. S. J. van der Zant, Phys. Rev. Lett. \textbf{96},
  026801 (2006).
\bibitem{hutt} A. K. H\"uttel, B. Witkamp, M. Leijnse,
  M. R. Wegewijs, and H. S. J. van der Zant,
  Phys. Rev. Lett. \textbf{102}, 225501 (2009).
\bibitem{leturq} R. Leturcq, C. Stampfer, K. Inderbitzin, L. Durrer,
  C. Hierold, E. Mariani, M. G. Schultz, F. von Oppen, and K. Ensslin,
  Nat. Phys. \textbf{5}, 327 (2009).
\bibitem{hutt3} A. K. H\"uttel, H. B. Meerwaldt, G. A. Steele,
  M. Poot, B. Witkamp, L. P. Kouwenhoven, and H. S. J. van der Zant,
  Phys. Status Solidi B \textbf{247}, 2974 (2010).
\bibitem{steele} G. A. Steele, A. K. H\"uttel, B. Witkamp, M. Poot,
  H. B. Merrwaldt, L. P. Kouwenhoven, and H. S. J. van der Zant,
  Science \textbf{325}, 1103 (2009).
\bibitem{lin} H. Lin, J. Lagoute, V. Repain, C. Chacon, Y. Girard,
  F. Ducastelle, H. Amara, A. Loiseau, P. Hermet, L. Henrard, and
  S. Rousset, Phys. Rev. B \textbf{81}, 235412 (2010).
\bibitem{lee} J. Lee, S. Eggert, H. Kim, S.-J. Kahng, H. Shinohara,
  and Y. Kuk, Phys. Rev. Lett. \textbf{93}, 166403 (2004).
\bibitem{chen} Y.-F. Chen, T. Dirks, G. Al-Zoubi, N. O. Birge, and
  N. Mason, Phys. Rev. Lett. \textbf{102}, 036804 (2009).
\bibitem{clauss} W. Clauss, D. J. Bergeron, M. Freitag, C. L. Kane,
  E. J. Mele, and A. T. Johnson, Europhys. Lett. \textbf{47}, 601
  (1999).
\bibitem{furu} M. Furuhashi and T. Komeda,
  Phys. Rev. Lett. \textbf{101}, 185503 (2008).
\bibitem{venema} L. C. Venema, J. W. G. Wild�er, J. W. Janssen,
  S. J. Tans, H. L. J. T. Tuinstra, L. P. Kouwenhoven, and C. Dekker,
  Science \textbf{283}, 52 (1999).
\bibitem{lemay} S. G. Lemay, J. W. Janssen, M. van den Hout, M. Mooij,
  M. J. Bronikowski, P. A. Willis, R. E. Smalley, L. P. Kouwenhoven,
  and C. Dekker, Nature (London) \textbf{412}, 617 (2001).
\bibitem{ouyang} M. Ouyang, J.-L. Huang, and C. M. Lieber,
  Phys. Rev. Lett. \textbf{88}, 066804 (2002).
\bibitem{buchs} G. Buchs, D. Bercioux, P.  Ruffieux, P. Gr\"oning,
  H. Grabert, and O. Gr\"oning, Phys. Rev. Lett. \textbf{102}, 245505
  (2009).
\bibitem{abedinpour} S. H. Abedinpour, M. Polini, G. Xianlong, and
  M. P. Tosi, Phys. Rev. A {\bf 75}, 015602 (2007).
\bibitem{qian} J. Qian, B. I. Halperin, and E. J. Heller, Phys. Rev. B
  {\bf 81}, 125323 (2010).
\bibitem{degio} U. De Giovannini, F. Cavaliere, R. Cenni, M. Sassetti,
  and B. Kramer, Phys. Rev. B {\bf 77}, 035325 (2008).
\bibitem{cavawi} F. Cavaliere, U. De Giovannini, M. Sassetti, and
  B. Kramer, New J. Phys. {\bf 11}, 123004 (2009).
\bibitem{giamarchi} T. Giamarchi, {\it Quantum Physics in One
  Dimension}, Oxford Science Publications (2004).
\bibitem{egger} R. Egger and A. O. Gogolin, Phys. Rev. Lett. {\bf 79},
  5082 (1997).
\bibitem{bercioux} D. Bercioux, G. Buchs, H. Grabert, and
  O. Gr\"oning, Phys. Rev. B \textbf{83}, 165439 (2011).
\bibitem{eggert} S. Eggert, Phys. Rev. Lett. \textbf{84}, 4413 (2000).
\bibitem{crep} A. Cr\'epieux, R. Guyon, P. Devillard, and T. Martin,
  Phys. Rev. B \textbf{67}, 205408 (2003).
\bibitem{leb} A. V. Lebedev, A. Cr\'epieux, and T. Martin,
  Phys. Rev. B \textbf{71}, 075416 (2005).
\bibitem{bena} C. Bena, Phys. Rev. B \textbf{82}, 035312 (2010).
\bibitem{suzu} H. Suzuura and T. Ando, Phys. Rev. B \textbf{65},
  235412 (2002).
\bibitem{mart} A. De Martino and R. Egger, Phys. Rev. B \textbf{67},
  235418 (2003).
\bibitem{mahan} G. D. Mahan, Phys. Rev. B \textbf{68}, 125409 (2003).
\bibitem{penn} G. Pennington and N. Goldsman, Phys. Rev. B
  \textbf{68}, 045426 (2003).
\bibitem{flens} K. Flensberg, New J. Phys. \textbf{8}, 5 (2006).
\bibitem{eros} E. Mariani and F. von Oppen, Phys. Rev. B \textbf{80},
  155411 (2009).
\bibitem{alves} M. Verissimo-Alves, R. B. Capaz, B. Koiller,
  E. Artacho, and H. Chacham, Phys. Rev. Lett. \textbf{86}, 3372
  (2001).
\bibitem{izu} W. Izumida and M. Grifoni, New J. Phys. \textbf{7}, 244
  (2005).
\bibitem{zazu} A. Zazunov, D. Feinberg, and T. Martin, Phys. Rev. B
  \textbf{73}, 115405 (2006).
\bibitem{shen} X. Y. Shen, B. Dong, X. L. Lei, and N. J. M. Horing, Phys. Rev. B
  \textbf{76}, 115308 (2007).
\bibitem{franck} J. Franck, Trans. Faraday Soc. \textbf{21}, 536
  (1926).
\bibitem{cond} E. Condon, Phys. Rev. \textbf{28}, 1182 (1926).
\bibitem{braig} S. Braig and K. Flensberg, Phys. Rev. B \textbf{68},
  205324 (2003).
\bibitem{oppen} J. Koch, F. von Oppen, and A. V. Andreev, Phys. Rev. B
  \textbf{74}, 205438 (2006).
\bibitem{koch} J. Koch and F. von Oppen, Phys. Rev. Lett. \textbf{94},
  206804 (2005).
\bibitem{haupt} F. Haupt, F. Cavaliere, R. Fazio, and M. Sassetti,
  Phys. Rev. B {\bf 74}, 205328 (2006).
\bibitem{merlo} M. Merlo, F. Haupt, F. Cavaliere, and M. Sassetti, New
  J. Phys. \textbf{10}, 023008 (2008).
\bibitem{piovano} F. Cavaliere, G. Piovano, M. Sassetti, and
  E. Paladino, New J. Phys. {\bf 10}, 115004 (2008).
\bibitem{cava} F. Cavaliere, E. Mariani, R. Leturcq, C. Stampfer, and M. Sassetti,
 Phys. Rev. B {\bf 81}, 201303(R) (2010).
\bibitem{yo} H. Yoshioka and Y. Okamura,
  J. Phys. Soc. Jpn. \textbf{71}, 2512 (2002).
\bibitem{grifoni} M. Grifoni and L. Mayrhofer, Eur. Phys. J. B {\bf
  56}, 107 (2007).
\bibitem{kleimann} T. Kleimann, F. Cavaliere, M. Sassetti, and
  B. Kramer, Phys. Rev. B {\bf 66}, 165311 (2002).
\bibitem{cavaprl} F. Cavaliere, A. Braggio, J. T. Stockburger,
  M. Sassetti, and B. Kramer, Phys. Rev. Lett. {\bf 93}, 036803
  (2004).
\bibitem{spineffects} F. Cavaliere, A. Braggio, M. Sassetti, and
  B. Kramer, Phys. Rev. B {\bf 70}, 125323 (2004).
\bibitem{jishi} R. A. Jishi, M. S. Dresselhaus, and G. Dresselhaus,
  Phys. Rev. B {\bf 48}, 11385 (1993).
\bibitem{bolotin} K. I. Bolotin, K. J. Sikes, J. Hone, H. L. Stormer,
  and P. Kim, Phys. Rev. Lett. {\bf 101}, 096802 (2008).
\bibitem{piovanoprb} G. Piovano, F. Cavaliere, E. Paladino, and
  M. Sassetti, Phys. Rev. B {\bf 83}, 245311 (2011).
\bibitem{ullersma} P. Ullersma, Physica (Amsterdam) {\bf 32}, 27
  (1966).
\bibitem{brandesnew} A. Metelmann and T. Brandes, arXiv:1107.3762 (2011).
\bibitem{yarnew}A. Donarini, A. Yar, and M. Grifoni, arXiv:1109.0723 (2011).
\end{thebibliography}
\end{document}